\documentclass[iop, revtex4]{emulateapj}
\usepackage{natbib}
\usepackage{array} 

\usepackage{graphicx}
\usepackage{latexsym}
\usepackage{wasysym}
\usepackage{amsfonts}

\usepackage{amsmath}
\usepackage{amssymb}
\usepackage{url}
\usepackage[utf8]{inputenc}

\usepackage{placeins}

\newcommand{\dadt}{\left\langle{da/dt}\right\rangle }
\newcommand{\shape}{{\tt shape\,}}
\newcommand{\degs}{^\text{$\circ$}}
\newcommand{\txtalign}[1]{\multicolumn{1}{c|}{#1}}
\newcommand{\aumy}{ au/My}

\begin{document}

\title{Asteroid 1566 Icarus's size, shape, orbit, and Yarkovsky drift from radar observations} \shorttitle{Asteroid 1566 Icarus}
\author{Adam H. Greenberg} \affil{University of California, Los Angeles, CA}
\author{Jean-Luc Margot} \affil{University of California, Los Angeles, CA}
\author{Ashok K. Verma} \affil{University of California, Los Angeles, CA}
\author{Patrick A. Taylor} \affil{Arecibo Observatory, HC3 Box 53995, Arecibo, PR 00612, USA}
\author{Shantanu P. Naidu} \affil{Jet Propulsion Laboratory, California Institute of Technology, Pasadena, CA}
\author{Marina. Brozovic} \affil{Jet Propulsion Laboratory, California Institute of Technology, Pasadena, CA}
\author{Lance A. M. Benner} \affil{Jet Propulsion Laboratory, California Institute of Technology, Pasadena, CA}

\begin{abstract}

Near-Earth asteroid (NEA) 1566 Icarus ($a=1.08$ au, $e=0.83$,
$i=22.8\degs$) made a close approach to Earth in June 2015 at 22 lunar
distances (LD). Its detection during the 1968 approach (16 LD) was the
first in the history of asteroid radar astronomy.  A subsequent
approach in 1996 (40 LD) did not yield radar images.
We describe analyses of our 2015 radar observations
of Icarus obtained at the Arecibo Observatory and the DSS-14 antenna at
Goldstone.
These data show that the asteroid
is a moderately flattened spheroid
with an equivalent
diameter of 1.44 km with 18\% uncertainties,
resolving long-standing questions about the asteroid size.  We also
solve for Icarus' spin axis orientation ($\lambda=270\degs\pm10\degs,
\beta=-81\degs\pm10\degs$), which is not consistent with the estimates
based on the 1968 lightcurve observations.  Icarus has a strongly
specular scattering behavior, among the highest ever measured in
asteroid radar observations, and a radar albedo of $\sim$2\%, among
the lowest ever measured in asteroid radar observations.  The low
cross-section suggests a high-porosity surface, presumably related to
Icarus' cratering, spin, and thermal histories.  Finally, we present
the first use of our orbit determination software for the generation
of observational ephemerides, and we demonstrate its ability to
determine subtle perturbations on NEA orbits by measuring Icarus'
orbit-averaged drift in semi-major axis ($(-4.62\pm0.48) \times
10^{-4}$\aumy, or $\sim$60 m per revolution).  Our Yarkovsky rate
measurement resolves a discrepancy between two published rates that
did not include the 2015 radar astrometry.

\end{abstract} 

\section{Introduction} \label{sec:intro}

Earth-based radar is a powerful tool for determining surface, subsurface, and
dynamical properties of objects within the Solar System. Radar imaging has been
used to obtain high-resolution images of planets, moons, asteroids, and comets,
as well as constrain their orbits and spin pole orientations. In particular,
radar is the only remote observing method, other than a physical
flyby mission, which can obtain multiple-aspect, sub-decameter shape
data for minor planets.

Radar is particularly useful for studying Near-Earth Objects (NEOs). As
fragments of
leftover planetesimals, NEOs serve as a window into the processes
that governed the formation of the Solar System. NEO studies probe the
accretional, collisional, erosional, radiative, and tidal processes
which shape the continued evolution of the minor
planets. Determination of NEO orbits and gravity environments opens
the door to human exploration of these objects, including potentially
valuable sample return missions.
Arguably most important of all, the discovery, categorization, and orbit
determination of NEOs, as mandated by the United States' Congress, helps
identify those objects which are potentially hazardous to life on Earth.

The first asteroid radar observations occurred at Haystack on June
13--15, 1968~\citep{pettengill1969} and Goldstone on June 14--16,
1968~\citep{goldstein1969}, during a 16
lunar distance (LD) close approach of the object 1566
Icarus. 
Icarus' highly
elliptical orbit ($a=1.08$ au, $e=0.83$, $i=22.8\degs$) has it passing within radar detection range of the
Earth only once every few decades.  In modern times, the radar
apparitions occurred in 1968, 1996, and 2015.  Following the 1996
radar observations, \citet{mahapatra1999} described Doppler spectra
but did not report a detection in
delay-Doppler
images. Here, we
report results from the 2015 Arecibo and Goldstone observations.

Although Icarus has only come within radar range of the Earth on three
occasions in modern times, the long temporal separation of these
events
is
invaluable for accurately constraining
the orbit of the asteroid, as well as measuring long term variations
in the orbit, such as those caused by the Yarkovsky effect. This is of
particular importance because of Icarus' status as a Potentially
Hazardous Asteroid (PHA),
although the current Earth Minimum Orbit Intersection Distance (MOID)
is 0.0344 astronomical units (au) or 13 LD.

In the following sections we
describe the first radar-derived
shape model for this object. We also discuss Icarus' unusual
radar scattering profile, along with potential explanations for its
scattering behavior, and compare these radar-derived results to
previous
thermal and
lightcurve results.
In addition, we present a new orbit determination program which
is capable of generating observational ephemerides for NEOs and
detecting the Yarkovsky effect for both radar and optically observed
objects.  We use this program to derive the magnitude of long-term
non-gravitational forces acting on Icarus.

\section{Observations of 1566 Icarus}

\begin{table*}
  \centering
\caption{ Pre-observational information generated for 1566 Icarus.
  Signal-to-noise ratio (SNR) calculations assume nominal transmitter
  power, but equipment problems reduced the available transmitter power
  at
  Arecibo Observatory
  (see text). } \setlength{\extrarowheight}{0.05cm}
\begin{tabular}{ |l|r|r|r|r|r|r|r| }
\hline \hline
 Date (UTC)		& RA (deg)	& Dec (deg)	& Distance (au)	& SNR/run	& Observatory	& Band \\ [1.5ex]
\cline {1-7}
\hline
 2015-06-13 13:14-06:01 & 106		& +63		& 0.072		&   8		& Goldstone	& X \\ [1.5ex]
 2015-06-14 14:34-07:58 & 137		& +64		& 0.061		&  13		& Goldstone	& X \\ [1.5ex]
 2015-06-16 20:00-09:06 & 190		& +42		& 0.054		&  22		& Goldstone	& X \\ [1.5ex]
 \hline
 2015-06-17 22:52-01:18 & 200		& +29		& 0.059		& 475		& Arecibo	& S \\ [1.5ex]
 2015-06-18 23:04-01:50 & 207		& +17		& 0.068		& 290		& Arecibo	& S \\ [1.5ex]
 2015-06-19 23:31-01:50 & 211		&  +8		& 0.080		& 165		& Arecibo	& S \\ [1.5ex]
 2015-06-21 00:08-01:30 & 215		&  +1		& 0.093		&  98 		& Arecibo	& S \\ [1.5ex]

\hline
\end{tabular}
\label{tbl:observations}
\end{table*}

\begin{table*}
\centering
\caption{Radar observations of asteroid 1566 Icarus. 
The first column indicates the telescope: Arecibo (A) or Goldstone (G).
UT Date is the universal-time date on which the observation began.
MJD is the corresponding modified Julian date.
Eph is the ephemeris solution number used.
RTT is the round-trip light time to the target.
P$_{\rm tx}$ is the transmitter power.
Baud is the length of each code element of the pseudo-random code used for imaging; it dictates the range resolution.
Spb is the number of complex samples per baud giving
a pixel height of baud/spb;
cw data are typically sampled at a rate of 12.5 kHz.
Res is the frequency resolution of the processed data.
Code is the length (in bauds) of the pseudo-random code used.
The timespan of the received data is listed by the UT start and stop times.
Runs is the number of completed transmit-receive cycles.}
\setlength{\extrarowheight}{0.05cm}
\begin{tabular}{rrrrrrrrrrrr}
\hline
\hline
Tel & UT Date & MJD & Eph & RTT & P$_{\rm tx}$ & Baud & Spb & Res & Code & Start-Stop & Runs \\
 & yyyy-mm-dd &  &  & s & kW & $\mu$s & & Hz &  & hhmmss-hhmmss & \\
\hline
G     & 2015 Jun 13 & 57186 & s82 & 70 & 411 & cw  &   &    2.0  & none  & 234028-235826 &  8\\
      &            &       & s82 &    & 412 & 10  & 1 &   24.6  & 127   & 000615-001914 &  6\\
      &            &       & s82 &    & 408 & 11  & 1 &   22.4  & 127   & 002354-004359 &  9\\
      &            &       & s84 &    & 416 & 11  & 1 &   22.4  & 127   & 004605-005642 &  5\\
      &            &       & s84 &    & 408 & 1   & 1 &    1.0  & 1023  & 010105-013038 & 13\\
  \\
\hline\\
G     & 2015 Jun 14 & 57187 & s90 & 61 & 410 & cw  &   &    0.5  & none  & 232137-234306 & 11\\
      &            &       & s90 &    & 410 & 10  & 1 &   24.6  & 127   & 234643-235146 &  3\\
      &            &       & s90 &    & 394 & 0.5 & 1 &    3.8  & 255   & 235736-001047 &  7\\
  \\
\hline\\
G     & 2015 Jun 16 & 57189 & s92 & 54 & 335 & 10  & 1 &   24.6  & 127   & 224941-225410 &  3\\
      &            &       & s92 &    & 397 & 0.5 & 1 &    0.5  & 255   & 230029-232821 & 16\\ 
      &            &       & s92 &    & 404 & cw  &   &    0.5  & none  & 233236-014530 & 73\\
      &            &       & s92 &    & 396 & 0.5 & 1 &    0.5  & 255   & 014958-032952 & 55\\
  \\
\hline\\
A     & 2015 Jun 17 & 57190 & u01 & 59 & 273 & cw  &   &    0.3  & none  & 233700-235711 & 7\\
      &            &       & u01 &    & 249 & 0.1 & 2 &    0.3  & 65535 & 000256-002248 & 10\\
      &            &       & u01 &    & 519 & 0.2 & 4 &    0.3  & 65535 & 004135-005943 & 6\\
  \\
\hline\\
A     & 2015 Jun 18 & 57191 & u01 & 67 & 596 & cw  &   &    0.4  & none & 232250-233726 & 7\\
      &            &       & u01  & 68 & 634 & 0.5 & 1 &    0.5  & 8191 & 234128-013824 & 52\\
  \\
\hline\\
A     & 2015 Jun 20 & 57193 & u01  & 80 & 622 & cw  &   &    1.0  & none & 002652-004924 & 9\\
      &            &       & u01  &    & 664 & 1   & 2 &    1.0  & 8191 & 005857-013453 & 14\\
  \\
\hline\\
A     & 2015 Jun 21 & 57194 & u01  & 93 & 642 & cw  &   &    1.5  & none & 004449-010517 & 7\\

\end{tabular}
\label{tbl:masterlog}
\end{table*}

Observations of Icarus were conducted from
2015 June 13 to June 16 at the Goldstone Observatory in California, and from
2015 June 17 to June 21 at the Arecibo Observatory in Puerto Rico.
During the Arecibo observations, the signal-to-noise ratio (SNR)
decreased by roughly a factor of 2 each day (Table~\ref{tbl:observations}).

At the Arecibo Observatory, continuous wave (CW) and delay-Doppler
data were taken at the S-band transmitter's nominal frequency of 2380
MHz. CW data were taken on each day of observations.  Delay-Doppler
data were taken on the first three days of observations. The baud
length of the delay-Doppler data, which controls the range resolution,
was adjusted each day as the Icarus--Earth distance increased, with
0.1 $\mu$s and 0.2 $\mu$s bauds (corresponding to 15 m and 30 m range
resolution, respectively) used on June 17, 0.5 $\mu$s (75 m) on June
18, and 1.0 $\mu$s (150 m) on June 19.  The CW data were reduced with
frequency resolutions ranging from 0.25 to 1.5 Hz, depending on that
day's SNR.  The delay-Doppler data were
reduced with frequency resolutions ranging from 0.30 Hz to 1 Hz.

During data-taking, transmission power was limited to $\sim$25\% of nominal (278
kW) on the
first day and $\sim$60\% (600 kW) on the remaining days, due to problems with the
transmitter klystrons (Table~\ref{tbl:masterlog}). This power limitation reduced the overall data quality
taken at Arecibo.

At Goldstone, CW and delay-Doppler data were taken
at the X-band transmitter's nominal frequency of 8560
MHz. Each day began with CW observations, followed
by coarse (10 or 11 $\mu$s baud) ranging and 1 $\mu$s or 0.5 $\mu$s
imaging.
The primary purpose of coarse ranging observations is to improve the
quality of an object's ephemeris.  The ranging measurements taken at
Goldstone were the first ever ranging observations of Icarus.  The CW
data were reduced with frequency resolutions ranging from 0.5 to 2 Hz,
depending on that day's SNR.  The delay-Doppler data were reduced with
frequency resolutions ranging from 0.50 Hz to 3.8 Hz.

\section{Methods} \label{sec:methods}

\subsection{Shape analysis}

\subsubsection{Radar scattering properties} \label{sec:scatter}
In order to determine a shape from radar images, one must
simultaneously fit for
the shape, spin, and radar scattering properties of the surface
and sub-surface of the object. Radio waves incident on the surface of
an object will
scatter over
a variety of angles with a range of associated
powers, and the relationship between scattering angle and
back-scattered power determines the object's specularity.
This scattering behavior can be described by $\frac{d\sigma}{dA}(\theta)$, or the
differential radar cross section per surface element area, as a function of
the incidence angle $\theta$~\citep{evan68}.

The \shape software package~\citep{hudson1994,magri2007} can estimate $\frac{d\sigma}{dA}$ by fitting for
the parameters of this function along with an object's surface shape.
We found that the Icarus data were well fit by either a two-component
cosine scattering law (Appendix~\ref{app:specularity}) or a `hagfors'
scattering law.
For the
analysis performed in this paper, we chose to use a `hagfors' scattering law, 
which is defined as
\begin{equation} \label{eq:scattering}
\frac{d\sigma}{dA} = H(\theta_0-|\theta|)\times\frac{RC}{2}(\text{cos}^{4}\theta+C\text{sin}^2\theta)^{-\frac{3}{2}}
\end{equation}
where $R$ and $C$ are tunable parameters of the scattering law,
$H(x)$ is the Heaviside step function, and $\theta_0$ is a fixed
angular cutoff value.  This scattering law is the standard Hagfors
Law~\citep{hagfors1964} with an angular cutoff.

\subsubsection{Shape and spin pole determination} \label{sec:gridmethod}
Shape analysis was performed using the \shape software
package with the fitting algorithm
described in \citet{greenberg2015}. 
The
shape was modeled with a triaxial ellipsoid during all stages of the
fitting process.
Attempts to utilize a more complex model (e.g., a spherical harmonic
model) resulted in no changes being applied during the minimization
process when starting from a best-fit ellipsoid, which suggests that
the data quality is not high enough to support a more complex
model. The spin period was fixed at $2.273$ hours
\citep{harris1998,gehrels1970}.

Our analysis used both continuous-wave and imaging data. Initial tests
showed that
including imaging data taken after the
first night of Arecibo observations (June 17)
had no effect on the fit.
This result is to be expected because the lower SNR
(Table~\ref{tbl:observations} and Figure~\ref{fig:cw}) for the later nights necessitated
relatively low-resolution images to be taken. Therefore, all
subsequent fits were performed only with imaging data taken at
$0.1\;\mu$s and $0.2\;\mu$s resolution on the first night of Arecibo
observations. Similarly, the Goldstone delay-Doppler images were not
included in the fits due to low resolution and SNR. All CW data from
both Arecibo and Goldstone were used.

The initial shape model was chosen as the ellipsoid representation of
a sphere with a radius of $0.63$ km~\citep{harris1998}.
We explored a variety of initial conditions for the spin pole using a
grid of evenly spaced positions, with neighboring points
$\sim$$15\degs$ apart.  During this grid search, the spin pole
positions were allowed to float along with the axial ratios of the
ellipsoid and scattering parameters.  Simultaneous fitting of both
size and spin pole parameters was made possible due to previous
modifications made to the \shape software~\citep{greenberg2015}.

We performed
a $\chi^2$ analysis to place constraints on the possible spin pole
orientations.  $\chi^2$ is the element-wise sum of squared differences
between a model's predicted CW power or delay-Doppler image pixel
value, and the corresponding measurement.  Specifically, after running
the grid search described above, we calculated the quantity
$\Delta\chi_i$ for each of the initial
grid positions, where
\begin{equation}
\Delta\chi_i \equiv \frac{\chi_i^2-\chi_\text{min}^2}{\sqrt{2k}},
\end{equation}
where $\chi_i^2$ is the $\chi^2$ associated with the $i^{\text{th}}$
grid position after convergence, $\chi_\text{min}^2$ is the lowest
$\chi^2$ achieved across all grid positions, and $k$ is the number of
degrees-of-freedom for the fit. 
$\Delta\chi_i$ can be thought of as
the distance in $\chi^2$-space between the $i^{\text{th}}$ grid
position and
that of the best fit, in units of standard
deviations of the $\chi^2$ distribution.
Grid positions with $\Delta\chi_i<1$
yield results that are statistically indistinguishable from
those of the best fit, 
and thus
represent positions with potentially valid shape, spin pole, and
scattering behavior solutions.

\subsection{Orbit Determination} \label{sec:idos}
Smearing of radar images in the range dimension occurs when the
object's range drift is not properly taken into account.  Provided
that the direction and magnitude of the target's motion is known prior
to taking data, this range drift can be
compensated for entirely at the data-taking stage by delaying the
sampling clock appropriately.  If there are residual errors due to
imperfect knowledge of the object's orbit, additional image
corrections at the pixel level are necessary before
summing individual images, which is suboptimal~\citep{ostro1993b}.
For this reason, radar observers place a high priority on securing an
accurate knowledge of the object's trajectory before the observations
or during the first few minutes of the observations.

Therefore, radar observations require the generation of ephemerides that encode
the knowledge of the object's position and velocity. Traditionally
we have used the On-site Orbit Determination Software
(OSOD)~\citep{giorgini2002} to generate ephemerides for radar observations.
For observations of Icarus, AHG, JLM, and AKV wrote a new
ephemeris-generating program called the Integration and Determination
of Orbits System (IDOS). Central to IDOS' operation is the Mission
analysis, Operations, and Navigation Toolkit Environment (MONTE), a
powerful tool developed by the Jet Propulsion Laboratory (JPL) for a
variety of space-related science and aeronautical
goals~\citep{evans16}.  MONTE has been used as an integral tool for
trajectory design and spacecraft tracking of most modern NASA
missions, and has therefore been thoroughly tested in this
context. More recently, MONTE has also been tested as a scientific
tool for mapping the internal structure of Mercury through
gravitational field determination~\citep{verma2016}.

IDOS utilizes MONTE's built-in orbital integrator, DIVA. DIVA uses
Adams's method with variable
timesteps to integrate the differential
equations of motion. DIVA can account for gravitational perturbations
from any set of masses, as well as arbitrary accelerations, including
non-gravitational forces.  For the analyses performed in this paper,
we considered the major planets and 24 of the most massive minor
planets~\citep{folkner2014} as gravitational perturbers.  During close
approaches, DIVA can take into account a detailed description of a
perturber's gravitational field. DIVA also considers general
relativistic effects during orbital integration.

IDOS also uses MONTE's built-in residual minimization
capabilities. The underlying algorithm is based on a Kalman
filter~\citep{kalman1960} and the work of \citet{bierman1977}.

IDOS can process both optical and radar astrometric
measurements. Optical measurements are automatically downloaded from
the Minor Planet Center (MPC) database~\citep{mpcdb}.  Measurements
are corrected for star catalog bias based on the reference catalog and sky
location, and weighted based on the reference catalog, observatory,
observation type, and date. Both debiasing and weighting are
calculated using the methods described in \citet{farnocchia2014}.

\subsection{Yarkovsky force model}

The Yarkovsky effect \citep[e.g.,][]{peterson1976,farinella1998,vokrouhlicky2000,rubincam2000}
is a secular thermal effect that causes an affected object's
semi-major axis to change over time, on the order of $10^{-4}$\aumy\ for a km-size object. By equating this thermal
acceleration $\ddot{r}$ with the change in momentum per unit mass due
to incident radiation (Appendix~\ref{app:yarkovsky}), one finds that
\begin{equation} \label{eq:ayark}
	\ddot{\vec{r}}= \xi \; \frac{3}{8\pi} \; \frac{1}{D\rho} \; \frac{L_{\odot}}{c} \; \frac{X_{\hat{p}}(\phi)\vec{r}(t)}{||\vec{r}(t)||^3},
\end{equation}
where $\vec{r}(t)$ is the heliocentric radial vector for the object at time $t$,
$\hat{p}$ is the unit spin-axis vector, $\phi$ is the phase lag,
$L_{\odot}$ is the luminosity of the sun, $c$ is the speed of light,
and $X_{\hat{p}}(\phi)$ is the rotation matrix about $\hat{p}$. $\xi$ is an
efficiency factor.

IDOS models the Yarkovsky effect by applying this acceleration at
every integration time step. The diameter $D$ and density $\rho$ are
assumed to be $1$ km and $1\text{ g}\text{ cm}^{-3}$ unless more
specific values can be determined for the object.  We note that these
assumptions do not ultimately have any effect on the final reported
value of $\dadt$ (Section~\ref{sec:yarkmsr}) ---any inaccuracies in
assumptions concerning $\rho$ and $D$ are absorbed by the efficiency
factor $\xi$.

Icarus's semi-major axis exhibits substantial variations as a function
of time due to close planetary encounters, including the 1968 and 2015
Earth encounters.  Our modeling of the Yarkovsky effect
considers all the gravitational dynamics but adds a non-gravitational
acceleration to the dynamical model.

\subsection{Yarkovsky determination}
\subsubsection{Measuring Yarkovsky drift}
\label{sec:yarkmsr}
As discussed in Section \ref{sec:idos}, we used our orbit
determination software to
detect and determine the magnitude of the Yarkovsky effect on Icarus.
Both positive and negative values for $\xi$ were considered, allowing
for the possibility that the object is either a retrograde or prograde
rotator, respectively.

We recorded a goodness-of-fit metric, $\chi^2$, where
\begin{equation}
\chi^2 = \sum_{i=0}^{N} \frac{(E_i - O_i)^2}{\sigma_i^2},
\end{equation}
where $E_i$, $O_i$ is the expected and observed value, respectively,
for the $i^{th}$ astrometric measurement, and $\sigma_i$ is the
measurement uncertainty for that observation, and $N$ is the number of
observations.
These measurements included both radar and optical astrometry.

We then identified the best-fit $\xi$ value and assigned one-standard-deviation error
bars corresponding to $\xi=\xi(\chi^2_{min}+1)$, as described in
~\citet{numericalrecipes}.

\subsubsection{Transverse acceleration to orbit-averaged drift} \label{sec:yarkanalytic}
As discussed in Section \ref{sec:yarkmsr},
our orbit determination software provides the capability to model the
magnitude of the instantaneous acceleration imparted on a rotating
object due to anisotropic reradiation of absorbed sunlight.  However,
for the purpose of comparing our results with those extant in the
literature, it is useful to convert between this quantity and the
orbit-averaged change in semi-major axis, $\dadt$. 

We note that because there is an arbitrary scaling factor ($\xi$) in our Yarkovsky force
model (Equation~\ref{eq:ayark}), we can select model parameters that
maximize the transverse component of the acceleration without
prejudice.  Therefore, we set the spin pole orientation parallel
to the orbital pole and the phase lag $\phi$ to $-90\degs$.  Neither
of these choices affect the final estimated value of $\dadt$
(Section~\ref{sec:yarkres}).

To find $\dadt$, we first considered the instantaneous change in semi-major
axis due to some small perturbing transverse force, as given by~\citep{burns1976}
\begin{equation}
\frac{da}{dt} =
\frac{2}{\sqrt{GM_{\odot}}}a^{\frac{3}{2}}(1-e^2)^{-\frac{1}{2}}(1+e\sin\theta_f)\frac{T}{m},
\end{equation}
where
$G$ is the gravitational constant, 
$M_{\odot}$ is the mass of the central body, $e$ and
$a$ are the orbital eccentricity and semi-major axis, respectively, $\theta_f$ is the
true anomaly at a specific epoch, $T$ is the perturbing transverse force, and $m$
is the mass of the orbiting body.

Substituting equation (\ref{eq:ayark}) for $\frac{T}{m}$, and
expressing $r(t)$ in terms of $\theta_f(t)$ via
\begin{equation}
r(\theta_f) = \frac{a(1-e^2)}{1+e\cos\theta_f},
\end{equation}
gives the instantaneous change in semi-major axis as a function of orbital
parameters and the Yarkovsky acceleration scaling parameter $\xi$

\begin{equation} \label{eq:burns}
\frac{da}{dt} = \xi \frac{3}{4\pi} \frac{L_{\odot}}{c\sqrt{GM_{\odot}}} \frac{1}{D\rho}  \frac{1}{\sqrt{a}}\frac{ (1+e\sin{\theta_f})(1+e\cos{\theta_f})^2}{(1-e^2)^{\frac{5}{2}}}.
\end{equation}

Averaging equation (\ref{eq:burns}) over one full orbit yields
\begin{equation}\label{eq:finalyark}
\dadt = \xi \frac{\hat\alpha}{\sqrt{a}} \frac{3L_{\odot}}{4\pi c\sqrt{GM_{\odot}}} \frac{1}{D\rho},
\end{equation}
where 
\begin{equation}
\hat\alpha = \frac{1}{2\pi}\int_{0}^{2\pi} \frac{ (1+e\sin{\theta_f})(1+e\cos{\theta_f})^2}{(1-e^2)^{\frac{5}{2}}} dM
\end{equation}
is a quantity that depends only on orbital parameters and the true
anomaly $\theta_f$ is a function of the mean anomaly $M$.

Noting that $\hat\alpha=1$ when $e=0$, we can plug nominal values into
equation (\ref{eq:finalyark}), yielding
\begin{equation}
  \begin{aligned}
    \dadt =  1.45 \; \times \; \hat\alpha & \left( \frac{1 \text{ au}}{a}\right)^{\frac{1}{2}} \left( \frac{\xi}{0.01} \right) \\
	& \left( \frac{1 \text{ km}}{D} \right) \left( \frac{ 1 \text{ g}\text{ cm}^{-3}}{\rho} \right)
    \times  \frac{10^{-4}\text{ au}}{\text{Myr}}.
  \end{aligned}
  \end{equation}

\subsubsection{Detection verification} \label{sec:verification}

Once a $\dadt$ has been estimated, it is necessary to verify whether
the measured Yarkovsky effect signal is sufficiently strong to be
considered a true detection. To determine this, we performed
an {\em analysis of variance} to compare our best-fit Yarkovsky model
($\xi=\xi_{b}$) to a model in which no Yarkovsky effect was present
($\xi=0$).

The analysis of variance was designed to compare the goodness-of-fit
between two models with different number of model parameters, a task
which is otherwise not straightforward.  We followed the methods
described in \citet{greenbook}.  Specifically, we calculated the
test-statistic
\begin{equation}
F = \frac{ \kappa_{\delta} }{ \kappa_{Y} }
\end{equation}
where 
\begin{equation}
\kappa_{\delta} =	\frac{ \sum_{i=0}^{N} (\frac{E_{0,i} - O_i}{\sigma_i})^2  -
			       \sum_{i=0}^{N} (\frac{E_{\xi_{b},i} - O_i}{\sigma_i})^2 }{m_Y-m_0} 
\end{equation}
and
\begin{equation}
\kappa_{Y} = \frac{\sum_{i=0}^{N} (\frac{E_{\xi_{b},i} - O_i}{\sigma_i})^2 }{N - m_Y}.
\end{equation}
Here, $E_{0,i}$ is the simulated $i^{th}$ observation assuming gravity
only ($\xi=0$), $E_{\xi_{b},i}$ is the simulated $i^{th}$ observation assuming a
Yarkovsky model with $\xi=\xi_{b}$, $O_i$ is the $i^{th}$ observation
and $\sigma_i$ is the measurement uncertainty for that observation,
$N$ is the number of observations, and $m_Y$, $m_0$ are the number of
free parameters in the Yarkovsky model ($m_Y=7$) and gravity-only
model ($m_0=6$), respectively.

We then calculate the value 
\begin{equation}
p = \int_{x=F}^{x=\infty} f_{{\rm (m_Y-m_0, N-m_Y)}}(x)dx ,
\end{equation}
where $f_{(m_Y-m_0, N-m_Y)}$ is the F-distribution probability density
function with $m_Y-m_0$ and $N-m_Y$ degrees of freedom.  The $p$-value
serves as a metric for testing the null hypothesis that the $\xi$
variable is superfluous.  In other words, a small $p$-value indicates
that it would be implausible for us to record the astrometry that was
actually observed in a gravity-only universe.  Small $p$-values suggest
that a non-gravitational component to the acceleration is required.
For instance, one could choose $p<0.003$ as a threshold to reject the
null hypothesis.
This criterion approximately corresponds to a 3-standard-deviation
detection.  We will see that we can reject the null hypothesis for
Icarus with much higher confidence than $p<0.003$.

\section{Results} \label{sec:results}

\subsection{Cross-section and polarization properties}

\begin{figure}[h!]
\includegraphics[width=250 pt]{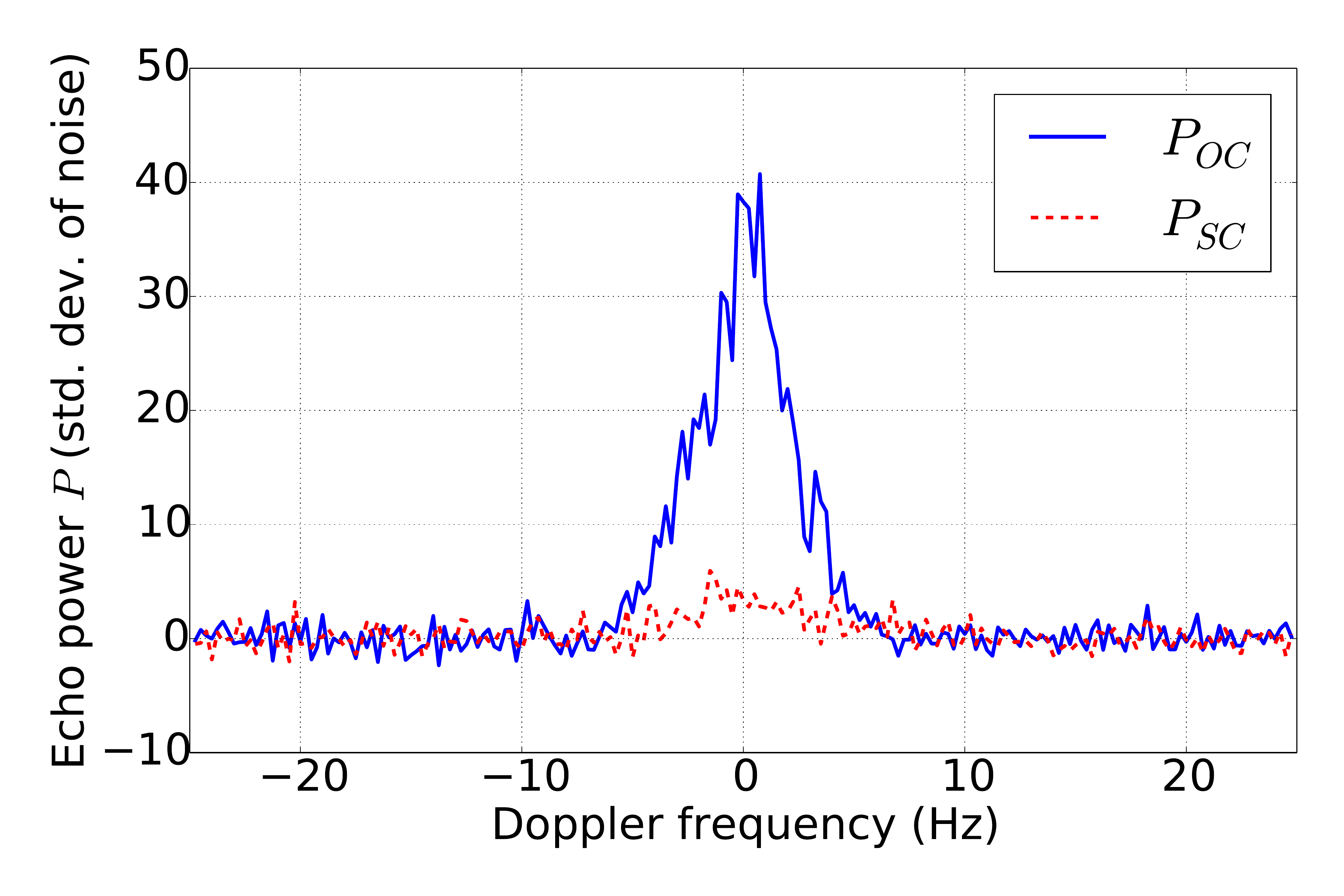}
\includegraphics[width=250 pt]{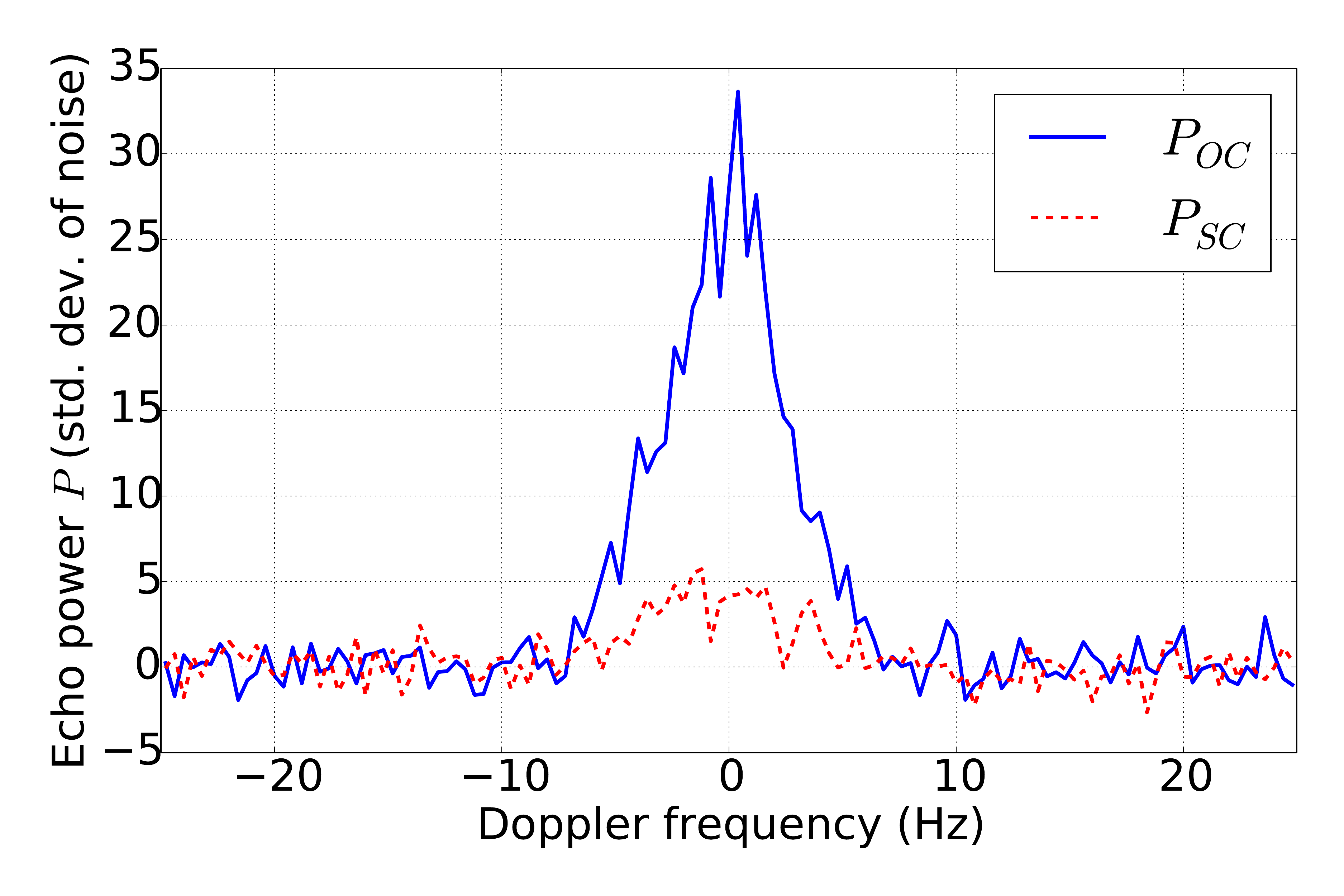}
\caption{
Echo spectra of 1566 Icarus obtained at Arecibo on 2015 June 17 at 0.25 Hz resolution
(top) and June 18 at 0.4 Hz resolution (bottom).
}
\label{fig:ocsc}
\end{figure}

The circular polarization ratio $\mu_{\rm C}$ is defined as the ratio
of the cross-section in the same sense circular polarization (SC) as
that transmitted to that in the opposite sense circular polarization
(OC). $\mu_{\rm C}$ serves as a measure of the surface roughness at
size scales comparable to the wavelength of the transmitted light.
The arithmetic average $\mu_{\rm C}$ for Icarus, as calculated from the Arecibo
(S-Band) CW observations, was $\mu_{\rm C}=0.18$ with a standard deviation of
$0.05$ (Figure~\ref{fig:ocsc} and Table~\ref{tbl:albedos}).
Some of the observed variations may be due to non-uniform scattering
properties over the asteroid's surface.

Icarus' total radar cross-section ($\sigma_{\rm T}$), defined as the sum of the
OC and SC cross-sections, was measured for each day of observations
using CW data.  
The arithmetic average $\sigma_{\rm T}$ as calculated from the Arecibo CW observations is
$0.030$ km$^2$ with a standard deviation of $0.007$ km$^2$.

The values for circular polarization ratio and total radar cross-section,
as calculated from the Goldstone (X-band) CW observations,
are $\mu_{\rm C}=0.33$ (standard deviation $0.05$) and
$\sigma_{\rm T}=0.069$ km$^2$ (standard deviation $0.041$ km$^2$).
\citet{mahapatra1999} had previously reported $\mu_{\rm C}=0.48\pm0.04$ at X-band.

\begin{figure*}[h!]
\centerline{ \includegraphics[width=550 pt]{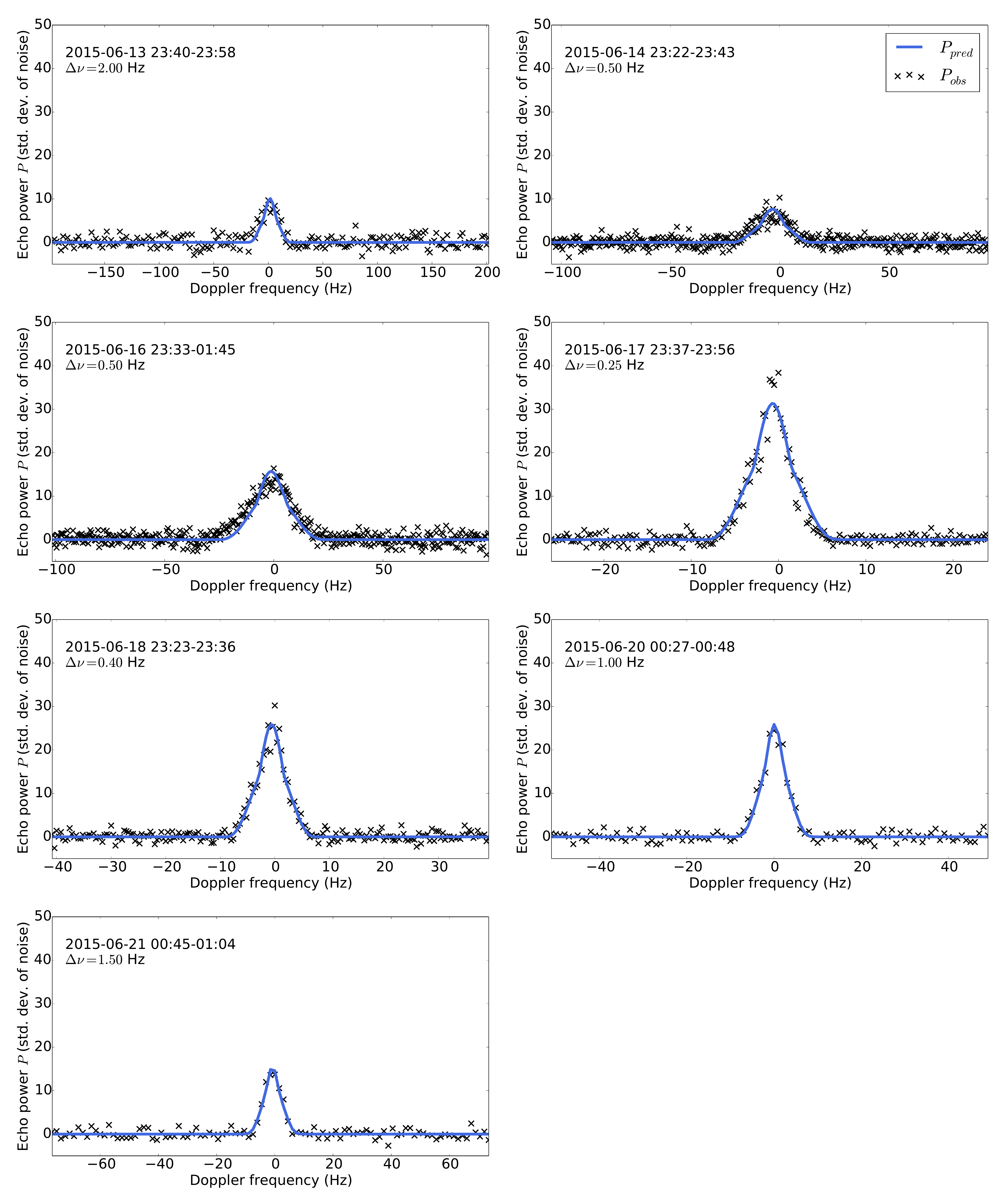} }
\caption{Continuous-wave observations of 1566 Icarus on each day of observation.
The first three sets of data were taken from Goldstone in the X-band, while the remaining four
sets were taken at Arecibo in the S-band. Each set represents the summed observed powers from
the corresponding day, compared to the summed predicted powers from our best-fit
model. $\Delta \nu$ is the frequency resolution.
}
\label{fig:cw}
\end{figure*}

There are two possible explanations for the discrepancies in
polarization ratios at S-band and X-band.  First, the low SNR of the
Goldstone observations may affect the determination of the
polarization ratios because receiver noise may inadvertently contribute
to the SC cross-section estimates, artificially raising the
$\mu_{\rm C}$ values.  For this reason, we believe that the Arecibo values are
more accurate due to the much higher SNR (Table~\ref{tbl:observations} or
Figure~\ref{fig:cw}).
Second, it is also possible that Icarus' cross-sections and
polarization ratios have a wavelength dependence. If so, the
difference between the Arecibo and Goldstone
$\mu_{\rm C}$ values would suggest that Icarus' surface is smoother at
S-band wavelength ($12.6$ cm) size scales than at X-band wavelength
($3.5$ cm).

\begin{table*}
  \centering
\caption{ Observed radar cross-sections ($\sigma$) and corresponding
  radar albedos ($\hat\sigma$) calculated on the basis of projected
  areas derived using our best-fit model.  The Arecibo S-band
  $\sigma_{\rm OC, SC}$ values have uncertainties of $\sim20\%$ due to
  uncertainties in the telescope gain and total power transmitted.
  Most of these absolute calibration errors cancel out when computing
  $\mu_{\rm C}$, for which we estimate uncertainties of $\sim 10\%$.
  One of us (SPN) calculated the Goldstone X-band $\sigma_{\rm OC,
    SC}$ and $\mu_{\rm C}$ values.  The Goldstone $\sigma_{\rm OC,
    SC}$ values have uncertainties of $35\%$ due to uncertainties in
  telescope pointing and other calibration errors, whereas the
  $\mu_{\rm C}$ uncertainty is $\sim 30\%$.  Latitude (lat) and
  longitude (lon) indicate the location of the sub-radar point, given
  in body-centric coordinates at the midpoint of the observation arc,
  with a prime meridian defined by the body's long axis.  The range of
  sub-radar point longitudes sampled over the duration of these
  observations, i.e., the rotational smear, is listed in the last
  column.  }

\setlength{\extrarowheight}{0.05cm}
\begin{tabular}{ |l|r|r|r|r|r|r|r|r|r|r|}
\hline \hline
Date (UTC)  & $\sigma_{\rm OC}$ (km$^2$)  &  $\sigma_{\rm SC}$ (km$^2$) & $\sigma_{\rm T}$ (km$^2$) & $\mu_{\rm C}$ & $\hat\sigma_{\rm T}$ & Band & lat (deg) & lon (deg) & rot. smear (deg)\\ [1.5ex] 
\cline {1-8}
\hline
2015 Jun 13 23:40-23:58 & 0.1151	& 0.0494	& 0.1645	& 0.429	& 0.111	& X & -40.5 & 258  & 48 \\ [1.5ex]
2015 Jun 14 23:22-23:43 & 0.0599	& 0.0192	& 0.0791	& 0.320	& 0.051	& X & -44.3 &  94  & 55 \\ [1.5ex]
2015 Jun 16 23:33-01:45 & 0.0361	& 0.0115	& 0.0476	& 0.318	& 0.031	& X & -41.3 & 191  & 348\\ [1.5ex]
\hline                                                                                               
2015 Jun 17 23:37-23:56 & 0.0364	& 0.0054	& 0.0418	& 0.149	& 0.028	& S & -33.5 &  81  & 50 \\ [1.5ex]
2015 Jun 18 23:23-23:36 & 0.0218	& 0.0042	& 0.0260	& 0.194	& 0.018	& S & -25.4 & 267  & 34 \\ [1.5ex]
2015 Jun 20 00:27-00:48 & 0.0201	& 0.0051	& 0.0252	& 0.256	& 0.018	& S & -18.5 & 246  & 55 \\ [1.5ex]
2015 Jun 21 00:45-01:04 & 0.0224	& 0.0031	& 0.0255	& 0.137	& 0.018	& S & -13.5 & 357  & 50 \\ [1.5ex]
\hline
\end{tabular}
\label{tbl:albedos}
\end{table*}

\subsection{Bandwidth}
\label{sec:bandwidth}
The observed bandwidth, as measured from the CW data, reached a minimum on the
second day of observations, which is consistent with the sub-radar latitude
reaching an extremum 
on that date (Table~\ref{tbl:bandwidths})
if the object is approximately spheroidal.

The predicted limb-to-limb bandwidth --- which can be calculated from the expression
\begin{equation}
B_{LL} = \frac{4 \pi D \cos{\delta}}{\lambda P},
\end{equation}
where $P$ is the object's rotational period, $\delta$ is the latitude of the
sub-radar point, and $\lambda$ is the wavelength of the transmitted signal ---
does not appear to agree with observations.
As we will discuss in detail in Section~\ref{sec:discussion},
this object's high specularity means that under certain observing conditions,
the echo signal may contain very little power from surface regions with high
incidence angles ($\theta_{inc}\gtrsim 45\degs$),
a fact already alluded to by \citet{pettengill1969} and \citet{goldstein1969}.
As a result, the calculated limb-to-limb bandwidth does not correspond to the
observed bandwidth, but the bandwidth measured at a level 1-standard-deviation above
the noise does correspond to the bandwidth calculated from the shape model at
the same signal strength.

\begin{table*}
  \centering
\caption{A comparison of the observed bandwidths at 1-standard-deviation above the noise ($B_{>1,O}$), to
  that predicted by our best-fit model ($B_{>1,P}$).
We attribute the discrepancy between the observed $B_{>1,O}$ and the predicted limb-to-limb
bandwidth ($B_{LL}$) to the highly specular surface of 1566 Icarus,
which, at low SNR, prevents the observer from receiving sufficient signal from
high-incidence-angle regions.
On the other hand, the $B_{>1,P}$ as predicted by the model matches the observed
$B_{>1,O}$, with the exception of June 16.
Bandwidth uncertainty was calculated as three times the frequency bin size.
All Goldstone bandwidths have been converted into S-band Hz for ease of comparison.
}
\setlength{\extrarowheight}{0.05cm}
\begin{tabular}{ |l|r|r|r|r| }
\hline \hline
& \multicolumn{1}{c|}{\textbf{Observed}} & \multicolumn{2}{c|}{\textbf{Predicted}}\\ [1.5ex]
\hline
Date (UTC) &  $B_{>1,O}$ (Hz) & $B_{LL}$ (Hz) & $B_{>1,P}$ (Hz) \\ [1.5ex]
\cline {1-4}
\hline
2015 Jun 13 23:40-23:58 & 10.4$\pm1.7$ & 12.7 &  9.5 \\ [1.5 ex]
2015 Jun 14 23:22-23:43 &  7.8$\pm0.4$ & 11.9 &  8.5 \\ [1.5 ex]
2015 Jun 16 23:33-01:45 & 12.8$\pm0.4$ & 14.2 & 10.2 \\ [1.5 ex]
\hline
2015 Jun 17 23:37-23:56 & 11.8$\pm0.8$ & 16.5 & 12.0 \\ [1.5 ex]
2015 Jun 18 23:23-23:36 & 13.3$\pm1.2$ & 18.1 & 12.8 \\ [1.5 ex]
2015 Jun 20 00:27-00:48 & 14.2$\pm0.8$ & 19.0 & 14.0 \\ [1.5 ex]
2015 Jun 21 00:45-01:04 & 14.5$\pm4.5$ & 19.4 & 15.0 \\ [1.5 ex]

\hline
\end{tabular}
\label{tbl:bandwidths}
\end{table*}

\subsection{Shape and spin pole determination} \label{sec:gridresult}
\begin{figure*}
\centerline{ \includegraphics[width=550 pt]{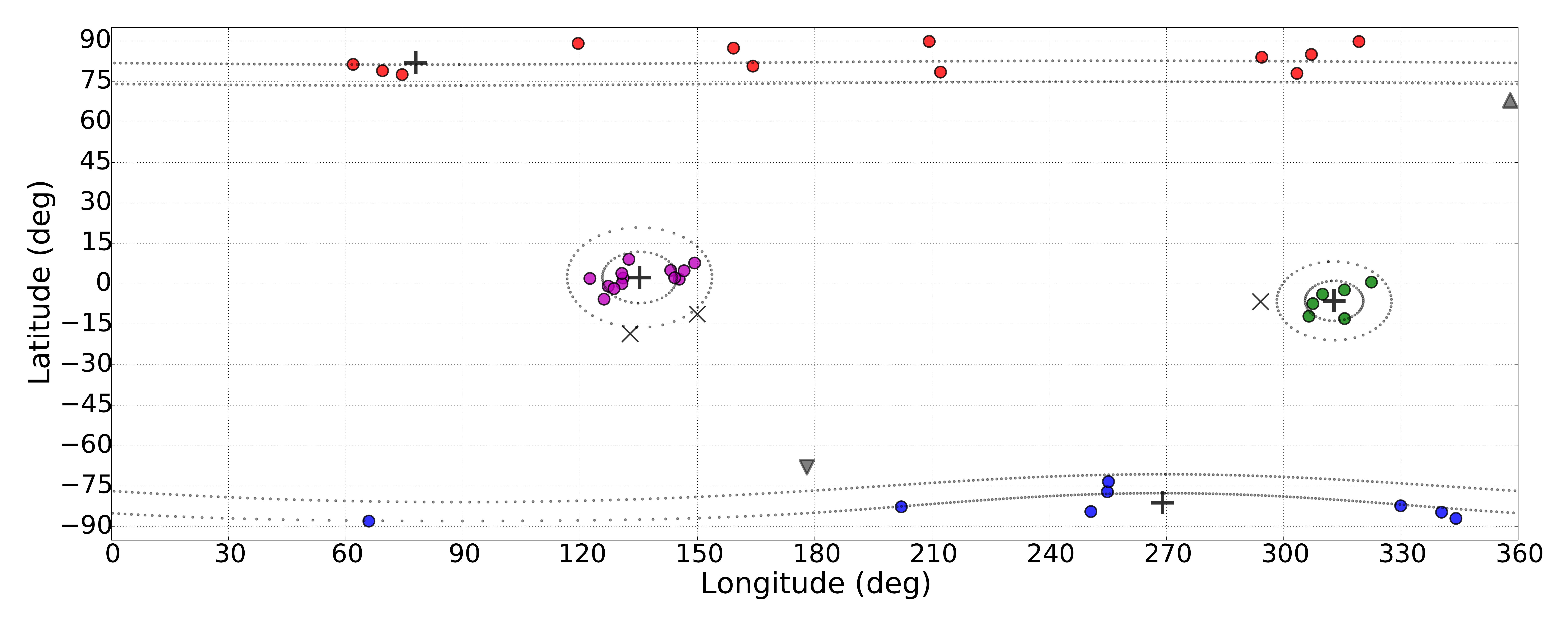} }
\caption{
  Results of our search for Icarus' spin axis orientation, in the ecliptic frame of J2000.
  We fit
  ellipsoids to the data with trial values of the spin axis
  orientation that were evenly distributed across the celestial
  sphere~\citep{deserno2004} with an initial spacing of $15\degs$.
  The spin pole location was allowed to float during these fits.
  Solutions with $\Delta\chi<\frac{1}{3}$ (Section~\ref{sec:gridmethod}) were plotted as circles and colored
  according to cluster membership. Black crosses indicate a cluster's
  mean position, whereas dotted lines show the 1- and
  2-standard-deviation uncertainty region in mean cluster
  position. Black x's indicate solutions that did not fall within any
  cluster with more than one member. Triangles show the location of
  the orbital poles.
}
\label{fig:clusters}
\end{figure*}

In Figure~\ref{fig:clusters}, we report solutions for the spin pole
orientations that satisfy $\Delta\chi_i\le\frac{1}{3}$
(Section~\ref{sec:gridmethod}).
After determining these solutions, we ran a clustering analysis based
on the nearest-neighbor algorithm~\citep{altman1992} to group
solutions with spatially correlated spin pole positions.
Figure~\ref{fig:clusters} shows cluster membership with coloring.  The
mean three-dimensional Cartesian position of these clusters, projected
onto the celestial sphere, are indicated with black crosses, and the
1- and 2-standard-deviation uncertainty regions in cluster mean position
are indicated with dashed circles. Note that projection effects make
these regions appear non-circular.

Table~\ref{tbl:solutions} shows the two possible pairs of spin pole positions, labeled by
their color (as seen in Figure~\ref{fig:clusters}). Each pair's member clusters (the purple and
green clusters, and the red and blue clusters) are
approximate antipodes of their partner.

The associated ellipsoid axis diameters and scattering parameters
(Section~\ref{sec:scatter}) are also listed in
Table~\ref{tbl:solutions}.  The uncertainty given for each parameter
was calculated as the standard deviation in said parameter among
members of the corresponding cluster.
Because the members of each cluster are not strictly independent
samples, these standard deviations may be underestimates of true
uncertainties, and do not account for potential systematic errors.

\begin{table*}
\centering
\caption{Possible spin poles (ecliptic coordinates) and corresponding size and
roughness values.  Our adopted solution (Section~\ref{sec:discussion}) is shown
in bold, 
and represents a moderately flattened spheroid.}

\setlength{\extrarowheight}{0.05cm}
\begin{tabular}{ |l|l|l|r|r|r|r|r| }
\hline \hline
\textbf{Cluster color} & \multicolumn{2}{c|}{\textbf{Spin pole}} & \textbf{Roughness} & \multicolumn{3}{c|}{\textbf{Ellipsoid axis diameters}} & {\textbf{Equiv. diameter}}\\ [1.5ex]
\cline {2-8}
{} &  \txtalign{\textbf{$\lambda$}} & \txtalign{\textbf{$\beta$}} & \txtalign{\textbf{$C$}} & \txtalign{\textbf{$2a$} (km)} & \txtalign{\textbf{$2b$} (km)} & \txtalign{\textbf{$2c$} (km) } & $D_{\rm eq}$ (km)\\ [1.01ex]
\hline
Purple & $135\degs\pm10\degs$ & $\;\;\;\;\;4\degs\pm10\degs$ & $12.9\pm2.1$ & $1.49\pm.15$ & $1.53\pm.16$ & $1.18\pm.13$ & $1.39\pm 10\%$\\ [1.5ex]
Green & $313\degs\pm10\degs$ & $\:\;-6\degs\pm10\degs$ & $13.9\pm0.7$ & $1.64\pm.12$ & $1.59\pm.12$ & $1.18\pm.27$ & $1.45\pm 13\%$\\ [1.5ex] 
\hline
\hline
Red & $78\degs\pm10\degs$ & $\:\;\;82\degs\pm10\degs$ & $13.3\pm2.0$ & $1.65\pm.19$ & $1.64\pm.19$ & $1.38\pm.20$ & $1.55\pm 13\%$\\ [1.5ex] 
\textbf{Blue} & $\mathbf{270\degs\pm10\degs}$ & $\mathbf{-81\degs\pm10\degs}$ & $\mathbf{13.8\pm1.4}$ & $\mathbf{1.61\pm.15}$ & $\mathbf{1.60\pm.17}$ & $\mathbf{1.17\pm.39}$ & $\mathbf{1.44\pm 18\%}$ \\ [1.5ex] 
\hline
\end{tabular}
\label{tbl:solutions}
\end{table*}

The parameters listed in Table~\ref{tbl:solutions} show the
specularity parameter $C$ (Equation~\ref{eq:scattering}) for the two pairs
of spin pole solutions. All solutions are consistent with a Hagfors
$C$ value of $13$.  The Hagfors formalism assumes that the surface
roughness has been smoothed with a wavelength-scale filter.  In this
formalism, the $C$ value corresponds to the root-mean-square slope
($S_0$) of a gently undulating surface at scales greater than the
wavelength.  From the expression
\begin{equation}
	S_0 = \frac{1}{\sqrt{C}},
\end{equation}
we find that the surface of Icarus has RMS slopes of $\sim$$16\degs$.
For comparison, ~\citet{evan68} found $S_0$ for the lunar maria of
$10.2\degs$ and $14.8\degs$ at wavelengths of $68$ cm and $3.6$ cm, respectively. 

\begin{figure*}[h!]
\centerline{ \includegraphics[width=500 pt]{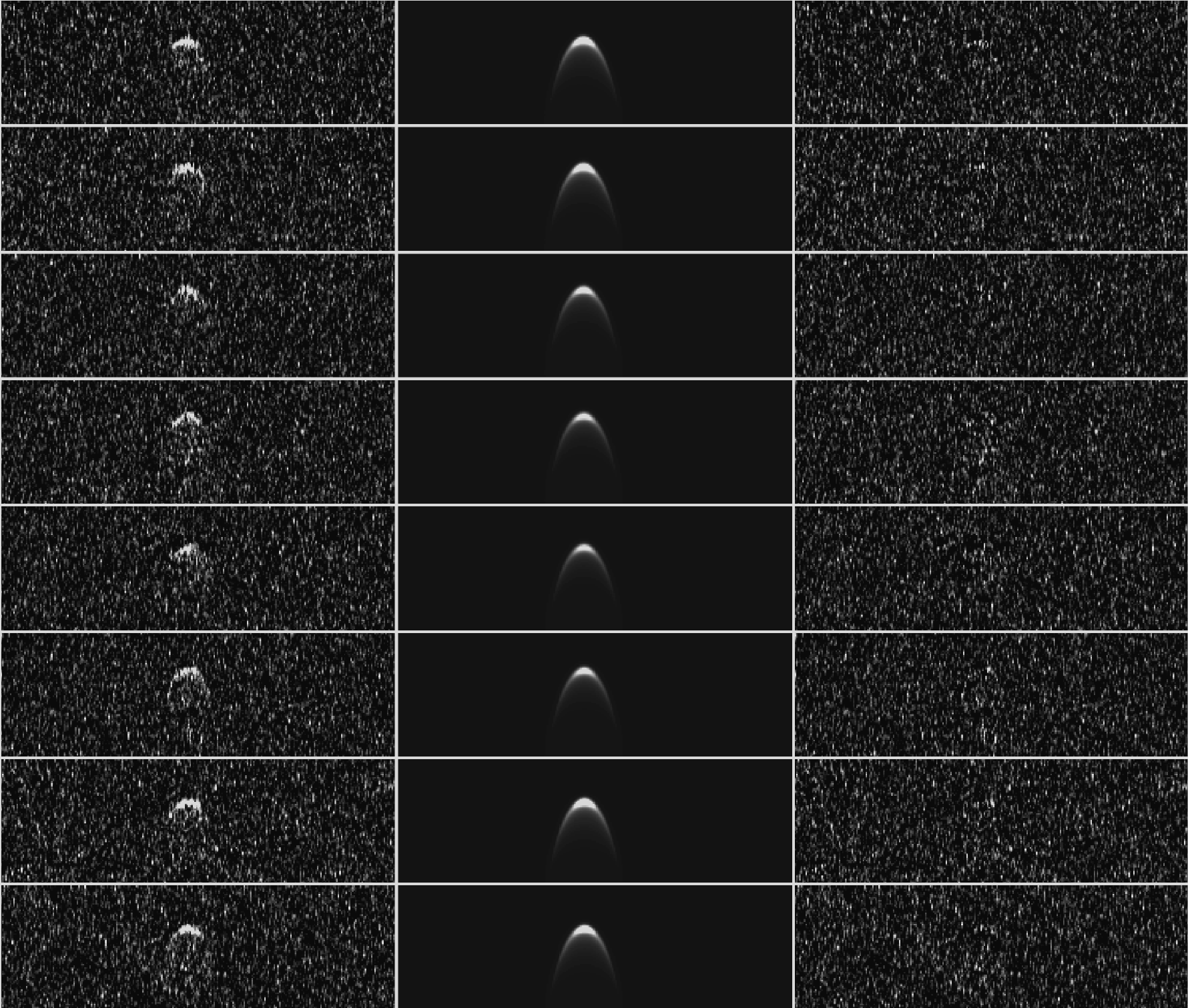} }
\caption{
  Delay-Doppler images of 1566 Icarus, showing (a) the observed data, (b)
  simulated images from the best-fit ellipsoid model, and (c) the residual
  images. The delay-Doppler observations of 1566 Icarus were taken from Arecibo
  Observatory on June 17.
  These data have a range resolution of $0.2\,\mu$s, corresponding to 30 m, with four
  samples per baud (i.e., each pixel represents 7.5 m), and reduced
  with a frequency resolution of $0.30$ Hz.  Each image includes $15$
  looks, or independent realizations.  Images are separated by $\sim$2
  minutes.  Within each image, range increases from top to bottom and
  frequency increases from left to right.  }
\label{fig:delaydop}
\end{figure*}

Figure~\ref{fig:delaydop} shows a series of delay-Doppler images
obtained on the first day of Arecibo observations, the corresponding
model images, and the residuals, while Figure~\ref{fig:cw} shows the
observed Doppler spectra compared to those predicted by our adopted model.
The simulated observations for this set of figures were generated from the
average shape, spin pole, and scattering parameters of
our adopted solution
(Table~\ref{tbl:solutions}, blue cluster).

As we will discuss in Section~\ref{sec:discussion}, this cluster’s spin pole is our preferred
solution because it is consistent with historical radar measurements and because
of the sign of Icarus’ semi-major axis drift.

\subsection{Astrometry and orbit refinement}
Our shape modeling allowed us to estimate the round-trip light times
between Arecibo's reference position and the center of mass of Icarus
with a fractional precision of $\sim 10^{-8}$.  We used the blue
cluster shape parameters for the purpose of computing radar astrometry
(Table~\ref{tbl:ranges}).

Despite a 60+ year arc of optical observations, the inclusion of the 2015
radar astrometry reduced uncertainties on orbital parameters by a factor of $\sim$3
(Table~\ref{tbl:uncertainties}).

\begin{table*}
\centering
\caption{
  Arecibo radar astrometry.  Round-trip light time
  measurements between Arecibo's reference position and the center of
  mass of 1566 Icarus at the receive times listed.  Estimates of the COM positions
  are based on fits of our best-fit shape model to the radar images.}

\setlength{\extrarowheight}{0.05cm}
\begin{tabular}{ |l|r|r| }
\hline
\hline
\cline {1-3}
Time (UTC) & RTT ($\mu$s) & 1-$\sigma$ uncertainty ($\mu$s) \\ [1.0 ex]
\hline
2015 Jun 18 00:02:00 & 58591220.06 & 0.4 \\ [1.5 ex]
2015 Jun 18 00:58:00 & 58866141.40 & 0.4 \\ [1.5 ex]
2015 Jun 18 23:41:00 & 67308377.46 & 1.0 \\ [1.5 ex]
2015 Jun 19 01:37:00 & 68163275.58 & 1.0 \\ [1.5 ex]
2015 Jun 20 00:58:00 & 79708409.31 & 2.0 \\ [1.5 ex]
2015 Jun 20 01:33:00 & 80022300.66 & 2.0 \\ [1.5 ex]

\hline
\end{tabular}
\label{tbl:ranges}
\end{table*}

\begin{table*}
  \centering
  \caption{
    Orbital elements and improvements in their formal
  uncertainties after inclusion of the 2015 radar astrometry.  The elements are
  semi-major axis $a$, eccentricity $e$, inclination $i$, longitude
  of the ascending node $\Omega$, argument of pericenter $\omega$, and
  mean anomaly $M$.
These orbital elements are valid at the epoch 2015 June 12 00:00:00 UT with our
nominal Yarkovsky drift solution (Section~\ref{sec:yarkres}). }
\setlength{\extrarowheight}{0.05cm}
\begin{tabular}{ |l|r|r|r|r| }
\hline \hline
\textbf{Parameter} & \textbf{Value} & \multicolumn{2}{c|}{\textbf{Uncertainty}}	& \textbf{Improvement factor}	\\ [1.5ex]
\cline {3-4}
		&	& Without 2015 radar data & With 2015 radar data	&			 	\\ [1.5ex]
\hline
$a$ (au)		&  1.077926624685	& 6.5e-11	& 2.6e-11 & 2.5		\\ [1.5ex]
$e$			&  0.826967321289	& 3.0e-08	& 6.9e-09 & 4.3		\\ [1.5ex]
$i$ (deg)		& 22.828097364019	& 6.9e-06	& 2.5e-06 & 2.8		\\ [1.5ex]
$\Omega$ (deg)		& 88.020929001348	& 1.6e-06	& 6.5e-07 & 2.5		\\ [1.5ex]
$\omega$ (deg)		& 31.363864782557	& 3.4e-06	& 1.2e-06 & 2.8		\\ [1.5ex]
$M$ (deg)		& 34.015936514108	& 1.7e-06	& 4.3e-07 & 4.0		\\ [1.5ex]

\hline
\end{tabular}
\label{tbl:uncertainties}
\end{table*}

\subsection{Yarkovsky drift}
\label{sec:yarkres}

We calculated the $\hat\alpha$ value for Icarus' orbit ($e=0.83$) of
$\hat\alpha=3.16$ (Equation~\ref{eq:finalyark}).
We then identified the best-fit efficiency factor $\xi$ (Section \ref{sec:yarkmsr}).
Assuming a density of
$\rho=2.7\text{ g}\text{ cm}^{-3}$ appropriate for Q-type asteroids
\citep{demeo2014} and our adopted effective diameter of $D=1.44\text{
  km}$, the best-fit efficiency factor is $\xi=(4.1\pm0.4)\%$.

The best-fit $\xi$ value corresponds to a semi-major axis drift rate
of $\dadt = (-4.62\pm0.48) \times10^{-4}$\aumy.
This lies within
one standard deviation of the value found by \citet{nuge12yark} of
$\dadt = (-3.2\pm2.0) \times10^{-4}$\aumy.  
\citet{farnocchia2013b} found a drift rate of $\dadt = (-0.86\pm1.8)
\times10^{-4}$\aumy.  Neither of these previous
results incorporated the 2015 radar astrometry, nor the more than 300 optical
observations taken of 1566 Icarus since 2012.

The $p$-value (Section~\ref{sec:verification}) for a Yarkovsky drift
model with $\dadt = (-4.62\pm0.48) \times10^{-4}$\aumy\ and 2312 degrees of
freedom is less than $10^{-10}$, leading to a confident rejection of the
null hypothesis (Section \ref{sec:verification}), which we interpret as a Yarkovsky detection.

In addition to conducting an analysis of variance (Table \ref{tbl:fscores}), we performed a variety of analyses to test the rigor of our
Yarkovsky result. A description of these tests can be found in
Appendix~\ref{app:rigorous}.

\begin{table*}
  \centering
\caption{
	Yarkovsky measurement results for 1566 Icarus.
        The observational arc for these measurements was 1949--2015. The
	radar astrometry includes both Arecibo and Goldstone measurements from 2015.
	$N_{\text{opt}}$ and $N_{\text{rad}}$ indicate the number of optical
	(after outlier rejection) and radar measurements, respectively.
        $F$ indicates the F-score, which serves as a measure of
	significance for the necessity of a  non-gravitational force component in the dynamical model (Section~\ref{sec:verification}). An
	F-score of 70 or above corresponds to a $p$-value of less than $10^{-15}$, or a detection at the $\geq 8 \sigma$ level.
	The rows labeled MPC indicate analysis done with the full MPC
	data, with basic outlier rejection (which discarded 80 optical
	observations), and observational
	weighting calculated using the methods described in~\citet{farnocchia2014}. 
	The rows labeled ``Screened'' indicate analysis done with a smaller data set
	from which astrometry deemed suspect on the basis of a gravity-only model was
	eliminated (Jon Giorgini, pers.\ comm.).
	A full description of the data and methods used can be found in Appendix~\ref{app:rigorous}.
	}
\setlength{\extrarowheight}{0.05cm}
\begin{tabular}{ |l|r|r|r|r|r|r|r| }
\hline \hline
 Data set used		& $N_{\text{opt}}$	& $N_{\text{rad}}$	& $\dadt$ & $F$ \\ [1.5ex]
\cline {1-4}
\hline
 MPC				& 1148		& 23		& $-4.6\pm0.5$ &   250	 \\ [1.5ex]
 				& 1148		& -		& $-4.9\pm0.5$ &   264	 \\ [1.5ex]
 \hline
Screened			& 931		& 23		& $-4.0\pm0.9$ &   139	 \\ [1.5ex]
 				& 931		& -		& $-3.8\pm1.1$ &   107	 \\ [1.5ex]
 \hline
Screened (JLM)                      & 931           & -             & $-3.6\pm1.0$ &    98	 \\ [1.5ex]
 \hline
\end{tabular}
\label{tbl:fscores}
\end{table*}

\begin{figure}[h!]
\includegraphics[width=250 pt]{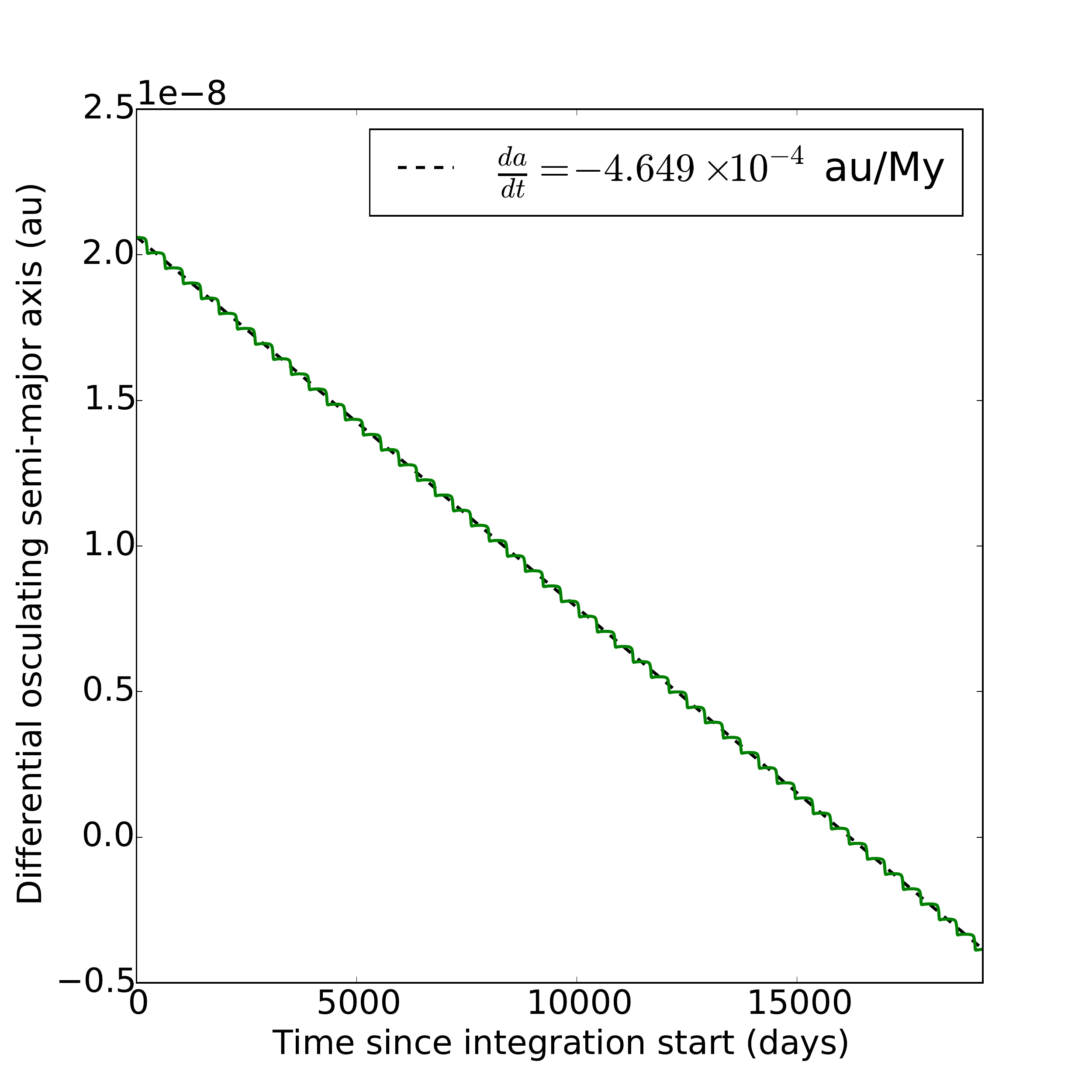}
\caption{ The difference in semi-major axis between a $\xi=0$
  Yarkovsky model and the best-fit $\xi$ value for a Yarkovsky model
  with spin pole located at ecliptic coordinate
  $\lambda = 270\degs$, $\beta = -81\degs$, 
  (the blue cluster).  The differential semi-major axis is plotted
  over the time interval 1963--2015.  The $\dadt$ has been numerically
  estimated with a linear fit through these data. The estimated drift in
  semi-major axis is consistent with the result obtained when assuming a spin
  pole that is parallel to the orbital pole,
  albeit with a different value of the adjustable parameter $\xi$ (Section~\ref{sec:yarkres}).}
\label{fig:drifts}
\end{figure}

We note that while our analysis of the thermal acceleration acting on
Icarus assumed a spin pole aligned with the orbital pole
(Section~\ref{sec:yarkmsr}), our final $\dadt$ value is insensitive to
this assumption.  We performed numerical estimates
(Figure~\ref{fig:drifts}) of the magnitude of the Yarkovsky effect for
the four possible spin pole orientations (Section~\ref{sec:gridresult}).
The drift was calculated as the change in position between the
best-fit Yarkovksy model and the best-fit model with no Yarkovksy
effect.  The $\dadt$ value shown was calculated as the slope of a
linear fit through these differences.  All spin pole orientations gave
consistent semi-major axis drifts. These drift rates are also
consistent with the rate of $(-4.62\pm0.48) \times10^{-4}$\aumy,
determined semi-analytically using equation (\ref{eq:finalyark}).

We plan on using our MONTE-based orbit determination software to
measure the orbital perturbations affecting other NEOs in Icarus-like
orbits~\citep{marg09iau261}. These measurements can in turn be used to
put constraints on the $\beta$ parameter in the post-Newtonian
parametrization of general relativity (GR) and possibly the oblateness
of the Sun.  The GR and Yarkovsky perturbations are essentially
orthogonal because Yarkovsky drift primarily affects the semi-major
axis, whereas GR does not affect the semi-major axis but instead
causes a precession of the perihelion.  The cleanest separation
between these perturbations will be obtained by solving for the orbits
of multiple asteroids simultaneously~\citep{marg09iau261}.

\section{Discussion} \label{sec:discussion}

\subsection{Spin pole}

\begin{table*}
\caption{ A comparison of measured bandwidths from previous radar observations
  of 1566 Icarus and bandwidths predicted for a variety of spin pole estimates.
  Antipodal pairs of spin poles yield bandwidths that
are identical within the margin of observing error, and thus only one of each
pair is shown.
 Table elements with an asterisk indicate a direct match
between that spin pole's bandwidth prediction and the corresponding bandwidth as reported by the authors.
Bold-faced table elements indicate a match between that spin pole's bandwidth
prediction and the estimated limb-to-limb bandwidth $B_{LL}$.
The $B_{LL}$ was estimated by adjusting 
for the decrease in apparent observed bandwidth caused by Icarus' highly specular surface, which
results in an approximate halving of the observed bandwidth as compared to the
nominal limb-to-limb bandwidth for very low SNR observations. The estimated $B_{LL}$ also accounts for the fact
that
some authors reported the half-max bandwidth, rather than zero-crossing bandwidth.
All bandwidth measurements and predictions have been converted to the
Arecibo S-band frequency of 2380 MHz for ease of comparison.  For observational details, see Appendix~\ref{app:historical}.}
\setlength{\extrarowheight}{0.05cm}
\begin{tabular}{ |l|r|r|r|r|r|r|r| }
\hline \hline
Observer & \multicolumn{2}{c}{{ \textbf{Pettengill et al.}}} & \multicolumn{1}{|c}{{ \textbf{Goldstein}}} & \multicolumn{1}{|c}{{ \textbf{Mahapatra et al.}}} & \multicolumn{3}{|c|}{{ \textbf{This work}}} \\ [1.5ex]
\hline
Year & \multicolumn{3}{c}{{1968}} &  \multicolumn{1}{|c}{{1996}} & \multicolumn{3}{|c|}{{2015}} \\ [1.5ex]
\hline
Date &  { June 13} & { June 14--15} & { June 15--16} & { June 8--10} & { June 14} & { June 18} & { June 21} \\ [1.5ex]
\hline 
\hline
Reported by authors	&  ${19.0}$ & ${4.0}$ & ${7.0}$ & 9.8 & ${10.4\pm1.7}$ & ${11.8\pm0.8}$ & ${14.5\pm4.5}$ \\ [1.5ex]
\hline
Estimated $B_{LL}$&  37.6 & 12.4 & 17.5 & 19.6 & 12.7 & 16.5 & 19.4 \\ [1.5ex]
\hline
\hline
 This work, green (predicted)          &  *17.4 & 18.5 & 19.4 & *12.1 & \textbf{14.2} & \textbf{16.6} & \textbf{19.6} \\ [1.5ex]
 This work, blue (predicted)		&   6.9	 & \textbf{12.0}& \textbf{16.3}  & \textbf{19.8} & \textbf{13.4} & \textbf{17.0} & \textbf{19.9} \\ [1.5ex]
 Gehrels et al. 1970 (predicted)	&  *19.9 & 20.0& 18.9           & \textbf{20.0} & 17.7 & 16.3 &  7.8 \\ [1.5ex]
 De Angelis 1995 (predicted)		&  *20.1 & 19.3& \textbf{17.5}           & \textbf{19.1}          & 19.5 & 12.9 &  3.4 \\ [1.5ex]
\hline
\end{tabular}
\label{tbl:npa}
\end{table*}

As noted in Section~\ref{sec:results}, analysis of our radar data
yields four possible sets of spin pole orientations for Icarus, and
given their low $\Delta\chi$ values, these solutions are statistically
indistinguishable from each other.  Icarus' spin pole has also been
measured in the past from lightcurve data, which gave three possible
spin poles.  Lightcurve observations of Icarus between 1968 June 14
and June 21 were analyzed by \citet{gehrels1970} and re-analyzed by
\citet{deAngelis1995}.  \citet{gehrels1970} reported a spin axis
orientation of $\beta = 0\degs\pm3\degs$, $\lambda = 223\degs\pm3\degs
\text{ or } 49\degs\pm3\degs$, whereas \citet{deAngelis1995} found
$\beta = 5\degs\pm5\degs$, $\lambda = 214\degs\pm5\degs$. These
lightcurve-derived spin poles are not consistent with our 2015 radar
data, which leaves two possibilities -- either the lightcurve-derived
spin poles are incorrect, or Icarus' spin pole has changed at some
point since the 1968 apparition.
We can determine which of these two possibilities is more likely by
analyzing historical radar measurements of this object.

Table~\ref{tbl:npa} lists Icarus' reported bandwidths from the three radar
apparitions that garnered CW measurements -- namely 1968 \citep{pettengill1969,
goldstein1969}, 1996 \citep{mahapatra1999}, and 2015 (this work) -- 
along with the bandwidths that would have been measured at the corresponding
dates, if Icarus' spin pole orientation were constant.
We generated predictions for four
possible spin poles -- $\lambda=313\degs\pm10\degs, \beta=-6\degs\pm10\degs$ (the
green cluster), $\lambda=270\degs\pm10\degs,\beta=-81\degs\pm10\degs$ (the blue
cluster), $\beta = 0\degs\pm3\degs, \lambda =
49\degs\pm3\degs$~\citep{gehrels1970}, and $\beta = 5\degs\pm5\degs,\lambda =
214\degs\pm5\degs$~\citep{deAngelis1995} -- and compared the predicted
measurements for these poles to the reported values. A direct comparison
suggests that no spin pole is consistent with measurements from more than one
apparition. However, this interpretation is slightly misleading, as there were a
variety of {\em types} of bandwidths reported
in these articles.
\citet{pettengill1969} and ~\citet{goldstein1969} reported bandwidths at full-width half-max
(FWHM), while \citet{mahapatra1999} reported their best estimate of the
limb-to-limb bandwidth (Appendix~\ref{app:historical}).
In this work, we reported the measured bandwidth
at one standard deviation above the noise ($B_{>1}$), as well as the limb-to-limb
bandwidth calculated from our shape models. Therefore, to facilitate comparisons
of these bandwidths, we have attempted to convert the historical measurements to
limb-to-limb bandwidths by estimating the signal's zero-crossing point, as well
as adjusting for the decrease in apparent bandwidth caused by Icarus' highly
specular surface (Section~\ref{sec:bandwidth}). When comparing the predicted
limb-to-limb bandwidths
with these measured limb-to-limb bandwidths, we find that our red/blue
cluster spin poles is consistent with all previous CW measurements of
Icarus, save the first spectrum obtained by \citet{pettengill1969} on
1968 June 13. However this measurement had a particularly low SNR, and
was considered potentially problematic by the observers. Therefore, if
we discard the first radar observation and assume that the
lightcurve-derived spin poles are incorrect, we can find a principal
axis rotation state (aligned with either the red or blue cluster
solutions) that is consistent with the historical data.
Furthermore, the slowly contracting orbit (Section~\ref{sec:yarkres})
is indicative of a spin pole anti-aligned with the orbital pole.
This suggests that, if Icarus is a principal axis rotator, it is likely to be a retrograde spinner, and thus aligned
with the blue cluster solution at ecliptic coordinates
\[\lambda=270\degs\pm10\degs,\]\[\beta=-81\degs\pm10\degs.\]

However, if the lightcurve-derived spin poles are correct, then the
spin axis orientation of Icarus would have had to have changed since
the 1968 apparition.  A difference in spin orientation may be caused
by non-principal axis (NPA) rotation or the effect of torques due to
anisotropic mass loss.  Mass loss may occur due to thermal fracturing.
NPA rotation may remain undetected in lightcurve or radar data if one
of the fundamental periods is long compared to the span of
observations in any given apparition or if the object is approximately
spheroidal.  NPA rotation requires a mechanism to excite the spin
state on a time scale shorter than the NPA damping time scale.  The
\citet{burn73} time scale for damping to principal axis rotation
assuming silicate rock ($\mu Q$ = 5 $\times 10^{12}$ Nm$^{-2}$,
$\rho=2.71\text{ g}\text{ cm}^{-3}$) is $\sim$25 million years, during
which Icarus experiences close planetary encounters that may excite
its spin.  However, if the material properties are closer to those
reported by \citet{sche15} ($\mu Q$ = 1.3 $\times 10^{7}$ Nm$^{-2}$),
NPA rotation would require an excitation in the past 100 years, which
we consider unlikely.

Three arguments favor the principal axis rotation solution: (1) the
lightcurve-derived spin poles predict bandwidths that are inconsistent
with observations (Table~\ref{tbl:npa}), (2) the bandwidths predicted
by the blue cluster spin pole solutions can match observations over
three separate apparitions, (3) damping to principal axis rotation
would likely occur on short time scales.  This conclusion suggests
that interpreting the evolution of lightcurve amplitudes may not be
sufficient in determining spin pole orientations accurately.  Other
considerations, such as knowledge of albedo variations and
inhomogeneous scattering behavior, may be necessary to properly
interpret lightcurve data.

\subsection{Cross-section and size}

Using the results of our analysis, we have found that 1566 Icarus'
radar scattering behavior is consistent with a Hagfors specularity constant of
$C=13$.  This level of specularity is unusual for NEOs
(Appendix~\ref{app:specularity}), and helps to explain the
lower-than-expected image quality. Diffusely-scattering surfaces can reflect
power from high incidence angles -- therefore, surface elements with normal
vectors not aligned with the observer's line of sight can still contribute to
the return signal. The surface elements of specular objects, on the other hand,
reflect most of their incident power away from the observer unless they lie
close to the sub-radar point (for ellipsoidal objects). The result is a very
sharp drop-off in SNR 
for incidence angles greater than $\sim$$20\degs$.
This effect can be seen in Figure~\ref{fig:delaydop}.

Furthermore, the small amount of echo power returning from regions
with high incidence angles ($\theta_{inc}\gtrsim 45\degs$) helps to
resolve the difference between Icarus' observed bandwidth and the
limb-to-limb bandwidth calculated from the object's shape and spin
pole orientation (Table~\ref{tbl:bandwidths}).  The most highly red-
and blue- shifted signals are reflected from these high-incidence
regions, and the large attenuation of these signals reduces the span
of the measured bandwidth.  This effect
may also explain a discrepancy noticed by
\citet{mahapatra1999}, who measured Icarus' bandwidth during its
previous radar apparition nearly two decades ago, and pointed out that the
measured bandwidth corresponds to a diameter around half of the value
found radiometrically by~\citet{harris1998}.

The results presented within this work suggest
Icarus' surface and sub-surface properties are
unusual
amongst the known population of NEOs. The radar albedo $\hat\sigma_T$
(defined as the radar cross-section divided by the object's geometric
cross-section) for an object with a diameter of $\sim$$1.44$ km and a
measured radar cross-section of $0.03 \text{ km}^2$ is less than 2\%
(see Table~\ref{tbl:albedos}), which we believe to be the lowest radar
albedo ever measured~\citep[e.g.,][]{magr99,magr01,magr07mbas}.

The radar albedo for a specular reflector provides an
approximation to the Fresnel reflectivity, which is related to the
dielectric constant~\citep{evan68}.  Our radar albedo measurement
yields the surprising low dielectric constant of 1.8.  Most
non-volcanic rocks have dielectric constants above 5, unless they are
in a powdered form with high porosity~\citep{camp69}, in which case
their dielectric constants are about 2.  Our measurements suggest that
Icarus may have a low surface density or, equivalently, a high
surface porosity.

Furthermore, a radar specularity of $C=13$ is
unusually high compared to most radar-imaged NEOs.
This high specularity and low radar albedo suggest an unusual
surface
structure.  Due to its highly eccentric orbit ($e=0.82$) and low
semi-major axis ($a=1.08$ au), Icarus approaches within $0.19$ au
of the Sun. At that distance, the equilibrium sub-solar point
temperature is expected to lie between $600$ K and $900$ K.
~\citet{jewitt2010} found that at temperatures within this range,
certain mineral compounds undergo extensive structural changes, possibly through the
mechanism of thermal fatigue \citep{delbo2014}.
It is possible that a combination of cratering history,
spin evolution, and repeated close approaches to the sun have modified
the surface of Icarus in such a way as to substantially lower its radar albedo.

Finally, we point out that the
equivalent diameter determined herein adds to the list of conflicting
sizes estimates for Icarus in the literature.  Using various
thermal models, \citet{harris1998} found a diameter of anywhere
between
$0.88$ km and $1.27$ km. \citet{mainzer2012} determined a
diameter of $1.36\pm0.43$ km using NEOWISE data at $3.4 \mu$m, $4.6
\mu$m, and $12 \mu$m, and the Near-Earth Asteroid Model (NEATM)
\citep{harris1998}.  Following their 1999 CW measurements of 
Icarus, \citet{mahapatra1999} calculated a diameter between $0.6-0.8$
km, assuming the spin pole reported by \citet{deAngelis1995}.  Finally, a
diameter can be calculated from the expression~\citep{fowler1992}
\begin{equation}
\label{eq:thermaldiam}
D = \frac{10^{-0.2H}1329}{\sqrt{p_V}},
\end{equation}
which, coupled with the measured H-magnitude of
$16.3$~\citep{harris1998} and geometric albedo $p_V$ of
0.14~\citep{thomas2011}, yields a diameter of $1.95$ km.  More recent
lightcurve measurements obtained at large phase angles yield $H=15.5$
\citep{mpbulletin}, which corresponds to a diameter of $2.80$ km.  We found
an equivalent diameter of 1.44 km with 18\% uncertainties.

\section{Conclusions}
In this work, we analyzed Arecibo and Goldstone radar observations of 1566 Icarus to
estimate its size, shape, scattering properties, orbital parameters,
and Yarkovsky drift. These results suggest that this object has
unusual surface properties, and
resolves long-standing questions about the object's size.

We presented the first use of our orbit-determination software and
demonstrated its ability to generate accurate radar ephemerides and to
determine the magnitude of subtle accelerations such as the Yarkovsky
effect.

\acknowledgements

AHG and JLM were funded in part by NASA grant NNX14AM95G and NSF grant
AST-1109772.  Part of the work done here was conducted at Arecibo
Observatory, which is operated by SRI International under a
cooperative agreement with the National Science Foundation
(AST-1100968) and in alliance with Ana G. Méndez-Universidad
Metropolitana (UMET), and the Universities Space Research Association
(USRA).  The Arecibo Planetary Radar Program is supported by the
National Aeronautics and Space Administration under Grant
Nos.\ NNX12AF24G and NNX13AQ46G issued through the Near-Earth Object
Observations program.  Some of this work was performed at the Jet
Propulsion Laboratory, which is operated by Caltech under contract
with NASA.  This work was enabled in part by the Mission Operations
and Navigation Toolkit Environment (MONTE).  MONTE is developed at the
Jet Propulsion Laboratory.  The material presented in this article
represents work supported in part by NASA under the Science Mission
Directorate Research and Analysis Programs.

\clearpage \FloatBarrier
\appendix
\section{A. Yarkovsky acceleration}
\label{app:yarkovsky}

For an object with diameter $D$, at distance from the Sun at time $t$ of  $r_t$,
the energy absorbed per second is
\[
\dot{E} = \frac{L_{\odot}}{4\pi r_t^2} \; \pi\left(\frac{D}{2}\right)^2,
\]
assuming perfect absorption.

The acceleration $\ddot{r}$ is equal to the photon momentum absorbed,
$\dot{p_{\gamma}}$, over the mass of the object, $m$.
\begin{align*}
\ddot{r} & =  \frac{\dot{p_{\gamma}}}{m} \\
 & = \frac{\dot{E}/c}{m} \\
 & = \frac{L_{\odot}}{4\pi r_t^2} \; \pi\left(\frac{D}{2}\right)^2 \; \frac{1}{c} \; \frac{1}{m}.
\end{align*}
Expressing the object mass in terms of density, $\rho$, and $D$, yields

\begin{align*}
\ddot{r} &  = \frac{L_{\odot}}{4\pi r_t^2} \; \pi\left(\frac{D}{2}\right)^2 \; \frac{1}{c} \; \frac{1}{\frac{4}{3}\pi\left(\frac{D}{2}\right)^3\rho} \\
 &  = \frac{3}{8\pi}\; \frac{L_{\odot}}{c} \; \frac{1}{r_t^2} \; \frac{1}{D\rho}.
\end{align*}

This acceleration is applied in the positive radial direction (in
heliocentric coordinates).
\begin{align*}
\ddot{\vec{r}} = \frac{3}{8\pi}\; \frac{1}{D\rho} \; \frac{L_{\odot}}{c} \; \frac{1}{r_t^2} \; \hat r
\end{align*}
Because it is purely radial, this acceleration will not cause a measurable
change in the orbit.

However, the absorbed photons will eventually be re-radiated, and induce an acceleration upon emission as
well. Since the object is rotating about some spin axis $\hat p$, this
secondary acceleration will not (necessarily) occur along a radial direction.
Furthermore, given that \textit{all} the absorbed photons must eventually be re-radiated, we
can express the magnitude of this secondary acceleration in the same manner as
its radial counterpart.  Here we define a phase lag, $\phi$, to describe at what
rotational phase (relative to the sub-solar longitude) the majority of the photons are re-emitted, and an efficiency
factor, $\xi$, which is
tied to the effective acceleration if one assumes that all photons are
re-emitted with phase lag $\phi$.
$X_{\hat{p}}(\phi)$ is the rotation matrix of angle $\phi$ about $\hat{p}$.
\[
	\ddot{\vec{r}}= \xi \; \frac{3}{8\pi} \; \frac{1}{D\rho} \;
	\frac{L_{\odot}}{c} \; \frac{X_{\hat{p}}(\phi)\vec{r_t}}{||\vec{r_t}||^3}.
\]

The Yarkovsky effect is caused by this \textit{non-radial} secondary acceleration.

\section{B. Historical data concerning 1566 Icarus' spin pole orientation}
\label{app:historical}

For the purpose of facilitating bandwidth comparisons, we normalize
all bandwidths to the Arecibo S-band frequency of 2380 MHz and label
the corresponding unit S-Hz. Except where otherwise noted, the
bandwidths discussed here are relayed as they were reported in the
corresponding
articles -- i.e., without any corrections applied for
specularity. We also note the distinction between reported half-power
bandwidths and zero-crossing bandwidths.

\subsection{Pettengill et al. 1968 radar observations}
The radar detection of Icarus by \citet{pettengill1969} on 1968 June
13 marked the first detection of an asteroid with radar.  These
observations were conducted at the Haystack Observatory at a frequency
of 7840 MHz.  A second set of observations took place on 1968 June 15.
\citet{pettengill1969} observed half-power Doppler
extents of 70 Hz on June 13 and 13 Hz on June 15.  Because both sets
of observations spanned multiple rotations of Icarus, the change in
bandwidth cannot be attributed to the shape of the object. 

The bandwidth reported on June 13 is over-estimated, for two reasons.
First, the authors reported a drift of their data-taking ephemeris by
8 Hz over the data-taking period, indicating that the echo was at most
62 Hz.  Second, the authors applied a variety of smoothing windows (8,
16, 32, and 64 Hz) and reported only the spectrum with the maximum
signal-to-noise ratio, which they obtained with the 64 Hz smoothing
window.  Because the convolution operation broadened the echo, we
estimate that the original, intrinsic echo was between 32 and 62 Hz.

Thus, the June 13 \citet{pettengill1969}
half-power bandwidth correspond to a value between 9.7 S-Hz and 19 S-Hz, and the
June 15 bandwidth corresponds to 4.0 S-Hz.

\subsection{Goldstein 1968 radar observations}
\citet{goldstein1969} detected Icarus between 1968 June 14 and 16
with a bistatic configuration at Goldstone.  The frequency was 2388
MHz and the reported half-power bandwidth for observations between June 15 04:30
UT and June 16 10:00 UT is about 7 Hz (or 7 S-Hz).

\subsection{Gehrels et al. 1968 lightcurve observations}
Lightcurve observations of Icarus between 1968 June 14 and June 21
were analyzed by \citet{gehrels1970} and re-analyzed by
\citet{deAngelis1995}.  \citet{gehrels1970} reported a spin axis
orientation of $\beta = 0\degs\pm3\degs$, $\lambda = 223\degs\pm3\degs
\text{ or } 49\degs\pm3\degs$, whereas \citet{deAngelis1995} found
$\beta = 5\degs\pm5\degs$, $\lambda = 214\degs\pm5\degs$. These spin poles do
not appear to be consistent with most of the radar CW bandwidths observed during the 1968,
1996, or 2015 apparitions (Table~\ref{tbl:npa}).

\subsection{Mahapatra et al. 1996 radar observations}
\citet{mahapatra1999} used Goldstone and observed a zero-crossing
Doppler bandwidth of 35 Hz at 8510 MHz.  Because they used a 10 Hz
smoothing window, the intrinsic bandwidth could be 35--45 Hz, or a
value between 9.8 and 12.6 S-Hz.

\section{C. Attempts to fit a lower specularity}
\label{app:specularity}

The specularity reported in this work for 1566 Icarus is unusually
high. When originally fitting the shape and scattering properties for
this object, we initially forced a lower specularity on our
models. After many such attempts, we came to the conclusion that a
diffusely-scattering surface did not match the data we observed.

\begin{figure*}[h!]
  \centering
\includegraphics[width=400 pt]{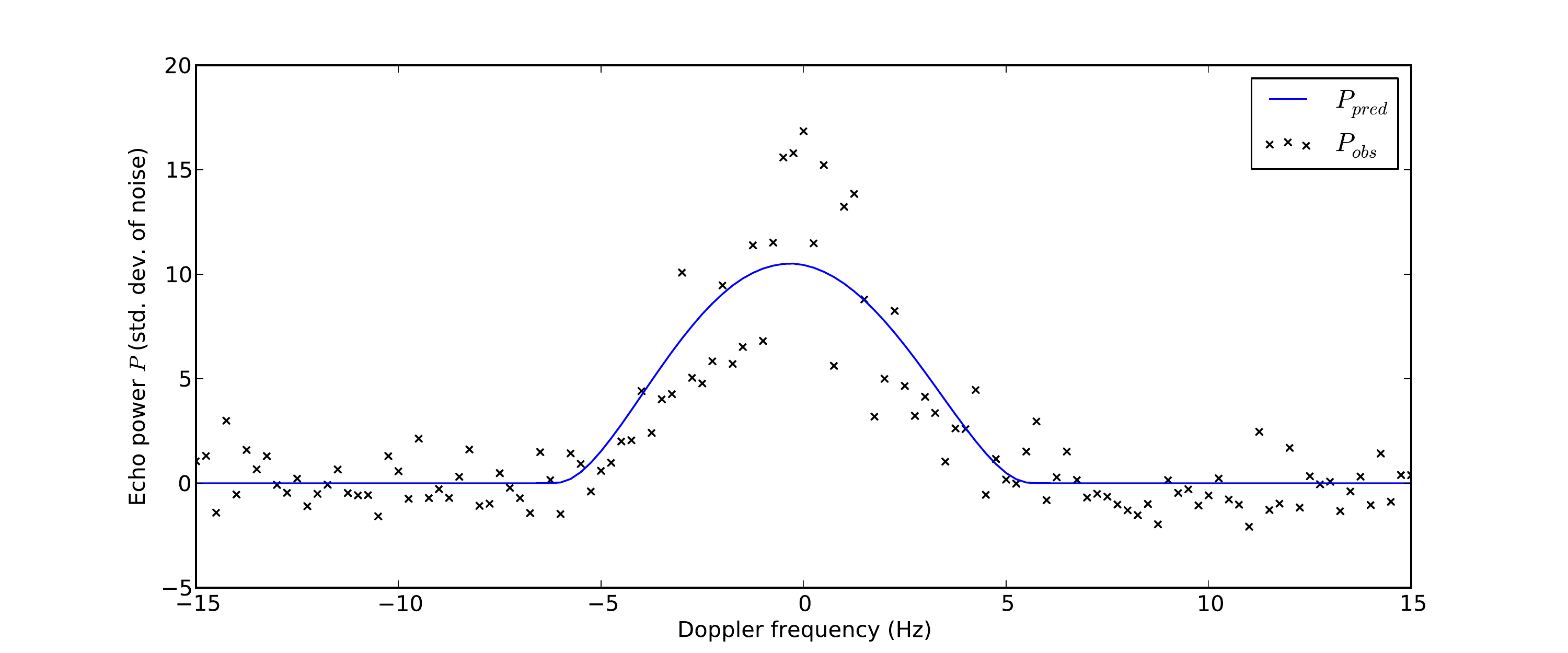} \\
\includegraphics[width=400 pt]{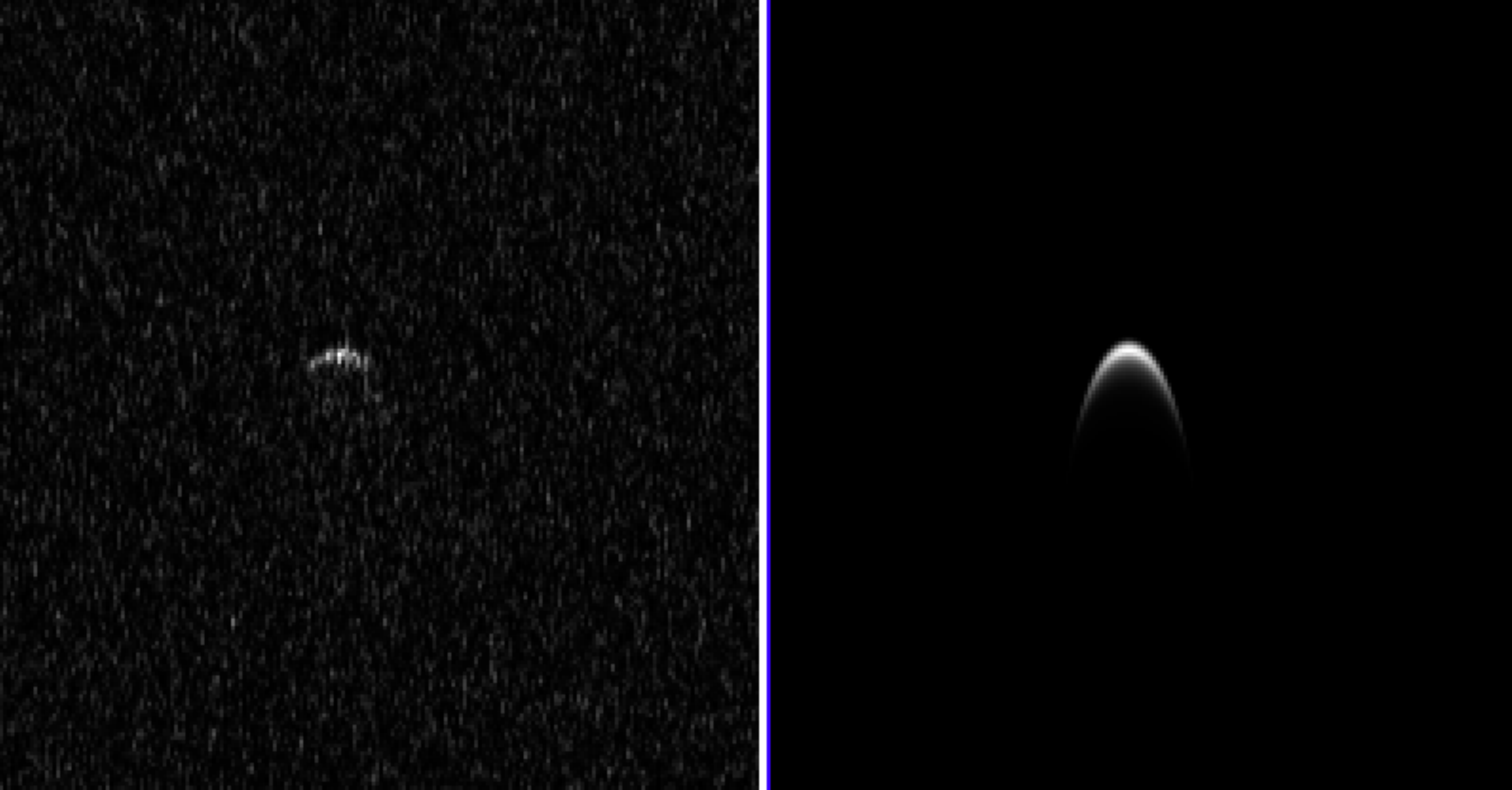}
\caption{
A result of a global fit to CW spectra and delay-Doppler images of 1566 Icarus,
with the specularity constant $C$ fixed to $2.0$. Shown is an example comparison between
the observed data and best-fit model spectrum for data taken on June 17,
(top), as well as a comparison between a best-fit model delay-Doppler image (bottom
right) and the corresponding observed image (bottom left), also taken on June
17. The fit shown does not match the data, and suggests that a higher specularity model is needed.
}
\label{fig:badfit}
\end{figure*}

Figure~\ref{fig:badfit} demonstrates what an attempted fit to Icarus'
CW spectra and delay-Doppler images looks like, when the specularity
is fixed to a lower value of $C=2$, and a `cos' scattering law is
utilized.
This scattering behavior can be defined with respect to the
differential radar cross section per surface element area
(Section~\ref{sec:scatter}), via
\begin{equation}
\frac{d\sigma}{dA} = R(C+1)\cos^{2C}{\theta}.
\end{equation}
These fits are the result of allowing all other model parameters
(ellipsoid axis ratios, signal scaling parameters, etc.) to float, and
performing a full fit on the same dataset used for the results
reported in this article. The best-fit model has an equivalent
diameter of $1.05$ km and, as the figure demonstrates, results in a
poor fit to the data. Note in particular that while the bounds of the
model spectrum is approximately equal to the bounds of the data
spectrum, the shape of the spectra do not match. In addition, the
signal in the delay-Doppler image does not drop off fast enough as
range from the observer increases. The fast drop-off noted in the data
necessitates a specular model.  Such a model can be realized with
either a two-component `cos' scattering law or a `hagfors' scattering
law.

\section{D. Additional evidence of Yarkovsky detection}
\label{app:rigorous}

We performed a variety of analyses to test the rigor of our Yarkovsky result
(Section~\ref{sec:yarkres}).

\subsection{Screening test}

We re-ran our Yarkovsky analysis with an independently curated set of
optical and radar astrometry (Jon Giorgini, pers.\ comm.). This data
set has been screened for potential outliers and faulty measurements
using a gravity-only model, with around 20\% of the original
astrometric data being discarded. Our analysis of these data (931
optical astrometric points and 23 radar astrometric points from 1949
to 2015) yielded a drift in semi-major axis of $\dadt=-4.0\pm0.9
\times10^{-4}$\aumy\ when including both radar and optical astrometry
and $\dadt=-3.8\pm1.1 \times10^{-4}$\aumy\ when including optical
astrometry only (Table~\ref{tbl:fscores}).

As a further verification, one of us (JLM) used the software he
developed in previous work \citep{nuge12yark} with this independently
screened data set and found $\dadt = (-3.6\pm1.0) \times10^{-4}$\aumy\ 
(Table~\ref{tbl:fscores}).

\subsection{Prediction test}

One way to analyze the accuracy of a Yarkovsky result is to check its
predictive power when compared to a
gravity-only
dynamical model.
Instead of collecting additional astrometry, which is not straightforward, 
we can simulate a prediction by re-analyzing a subset of the data
before some fiducial point in time, $t_f$, and then checking how the
resulting trajectory fares at predicting observations that were taken
after $t_f$.

The first range measurement of Icarus was obtained from Goldstone on
June 14, 2015 (Table~\ref{tbl:masterlog}). We therefore chose
2015-June-13 23:50 UT as $t_f$. We fit a Yarkovsky model to the data
taken before $t_f$ (936 optical observations and 11 radar
observations, from which 55 optical observations were discarded as
outliers), and found a semi-major axis drift rate of $\dadt =
(-3.7\pm0.7) \times10^{-4}$\aumy. This fit yielded a goodness-of-fit
of $\chi^2=634$. We also fit a gravity-only model to the same set of
data, which resulted in a goodness-of-fit of $\chi^2=662$. We then
compared how well these best-fit trajectories could predict the first
Icarus range measurement on June 14, 2015. The best-fit Yarkovsky
model residual for this prediction was 97 km, while the best-fit
gravity-only model residual was 267 km. This demonstrates that the
Yarkovsky model more accurately predicted a future measurement than
the gravity-only model.

\subsection{Incorporation test}

Another test of rigor is an incorporation test. After performing the
prediction test described above, we then added the first Doppler and
range measurements of 2015
and fit both a gravity-only model and a Yarkovsky model once again.
No additional outlier rejection was allowed.
With these  new Doppler and range measurements, the best-fit semi-major axis drift
rate for the Yarkovsky model is $\dadt = (-4.7\pm0.5)
\times10^{-4}$\aumy. The goodness-of-fit for the Yarkovsky model is
$\chi^2=638$, or a $<1$\% increase as compared to the Yarkovsky fit
prior to including the new
radar measurements. The best-fit
gravity-only model yielded a goodness-of-fit of $\chi^2=719$, or a 9\%
increase as compared to the gravity-only fit prior to including the
new radar measurements. These results suggest that incorporating the
first Icarus radar measurements into a gravity-only model results in a
general degradation in the quality of the fit to optical
observations. However, these same radar measurements can be included in
a Yarkovsky model with no appreciable effect on the goodness-of-fit.

\subsection{Combined test}
Finally, we re-ran both the prediction test and the incorporation test
described above on the curated data set.
The best-fit Yarkovsky model fit to data taken before $t_f$ yielded a
drift in semi-major axis of $\dadt = (-3.53\pm
1.11)\times10^{-4}$\aumy, and a goodness-of-fit of $\chi^2=172$, while
the gravity-only model had a goodness-of-fit of $\chi^2=182$. The
Yarkovsky model predicted the first Goldstone range measurement with a
residual of 43 km, while the gravity-only model predicted residual for
that data point was 128 km.  For the incorporation test, we again
added the first Goldstone Doppler and range measurements to the data set. For the
Yarkovsky model, the best-fit drift in semi-major axis was $\dadt =
(-3.95\pm 0.97)\times10^{-4}$\aumy, with a goodness-of-fit of $\chi^2=173$,
or an increase of $<1$\% as compared to the Yarkovsky fit prior to
adding the new radar measurements.  The best-fit gravity-only model had
a goodness-of-fit of $\chi^2=189$, or a $4$\% increase as compared to
the gravity-only model fit prior to adding the new radar
measurements. Note that these fits were performed on a data set for
which outlier rejection had been performed assuming a gravity-only
model --- even so, a gravity-only model still required a marked decrease in fit
quality in order to incorporate the first radar measurements, while the
Yarkovsky model saw no appreciable change in goodness-of-fit.

\subsection{Summary}

The tests that we performed are summarized in
Table~\ref{tbl:prediction} and Table~\ref{tbl:incorporation}, and
confirm the robustness of the Yarkovsky detection.

\begin{table*}
  \centering
\caption{ The residuals of the first range measurement obtained of 1566 Icarus
when using a Yarkovsky model and a gravity-only model, and when fitting these
models only to data taken before the first radar measurements of 2015 (i.e., the
prediction test). }
\setlength{\extrarowheight}{0.05cm}
\begin{tabular}{ |l|r|r| }
\hline \hline
& \multicolumn{2}{c|}{Residual (km)} \\ [1.5ex]
\hline
Data set	& Yarkovsky model & Gravity-only model	\\ [1.5ex]
\hline
MPC		& 97			& 267		\\ [1.5ex]
Screened	& 43			& 128		\\ [1.5ex]

\hline
\end{tabular}
\label{tbl:prediction}
\end{table*}

\begin{table*}
  \centering
\caption{ The goodness-of-fit ($\chi^2$) for a Yarkovsky model and a
gravity-only model when fit to only data taken before the first radar 
measurements of 2015 (i.e., before $t_f=$ 2015-June-13 23:50:00 UT),
and how $\chi^2$ changes when the first Doppler and range measurements are
incorporated into the fit (i.e., the incorporation test).}
\setlength{\extrarowheight}{0.05cm}
\begin{tabular}{ |l|r|r| }
\hline \hline
\multicolumn{3}{|c|}{\textbf{MPC data set}}\\ [1.5ex]
\cline {1-3}
					& Yarkovsky model	& Gravity-only model		\\ [1.5ex]
All data prior to $t_f$                 & 634			& 661				\\ [1.5ex]
All data prior to $t_f$ + 1 range, 1 Doppler	& 638			& 719				\\ [1.5ex]
\hline
Fractional increase in $\chi^2$		& $<1$\%		& 9\%				\\ [1.5ex]
\hline
\hline
\multicolumn{3}{|c|}{\textbf{Screened data set}}\\ [1.5ex]
\cline {1-3}
					& Yarkovsky model	& Gravity-only model		\\ [1.5ex]
All data prior to $t_f$                 & 172			& 182				\\ [1.5ex]
All data prior to $t_f$ + 1 range, 1 Doppler	& 173			& 189				\\ [1.5ex]
\hline
Fractional increase in $\chi^2$		& $<1$\%		& 4\%				\\ [1.5ex]

\hline
\end{tabular}
\label{tbl:incorporation}
\end{table*}

\newpage \FloatBarrier

\clearpage
\bibliography{comps_bib}

\begin{thebibliography}{}
\expandafter\ifx\csname natexlab\endcsname\relax\def\natexlab#1{#1}\fi

\bibitem[{Altman(1992)}]{altman1992}
Altman, N.~S. 1992, The American Statistician, 46, 175

\bibitem[{Bierman(1977)}]{bierman1977}
Bierman, G.~J. 1977, "{Factorization Methods for Discrete Sequential
  Estimation}" ({Dover})

\bibitem[{{Burns}(1976)}]{burns1976}
{Burns}, J.~A. 1976, American Journal of Physics, 44, 944

\bibitem[{{Burns} \& {Safronov}(1973)}]{burn73}
{Burns}, J.~A., \& {Safronov}, V.~S. 1973, \mnras, 165, 403

\bibitem[{Campbell \& Ulrichs(1969)}]{camp69}
Campbell, M.~J., \& Ulrichs, J. 1969, J. Geophys. Res., 74, 5867

\bibitem[{{De Angelis}(1995)}]{deAngelis1995}
{De Angelis}, G. 1995, Planet. Space Sci., 43, 649

\bibitem[{{Delbo} {et~al.}(2014){Delbo}, {Libourel}, {Wilkerson}, {Murdoch},
  {Michel}, {Ramesh}, {Ganino}, {Verati}, \& {Marchi}}]{delbo2014}
{Delbo}, M., {Libourel}, G., {Wilkerson}, J., {et~al.} 2014, \nat, 508, 233

\bibitem[{{DeMeo} {et~al.}(2014){DeMeo}, {Binzel}, \& {Lockhart}}]{demeo2014}
{DeMeo}, F.~E., {Binzel}, R.~P., \& {Lockhart}, M. 2014, Icarus, 227, 112

\bibitem[{{Deserno}(2004)}]{deserno2004}
{Deserno}, M. 2004, {How to generate equidistributed points on the surface of a
  sphere}, \url{https://www.cmu.edu/biolphys/deserno/pdf/sphere\_equi.pdf}

\bibitem[{Evans \& Hagfors(1968)}]{evan68}
Evans, J.~V., \& Hagfors, T., eds. 1968, Radar Astronomy (New York:
  McGraw-Hill)

\bibitem[{{Evans} {et~al.}(2016){Evans}, {Taber}, {Drain}, {Smith}, {Wu},
  {Guevara}, {Sunseri}, \& {Evans}}]{evans16}
{Evans}, S., {Taber}, W., {Drain}, T., {et~al.} 2016, in The 6th International
  Conference on Astrodynamics Tools and Techniques (ICATT), International
  Conference on Astrodynamics Tools and Techniques, Darmstadt, Germany

\bibitem[{Farnocchia {et~al.}(2015)Farnocchia, Chesley, Chamberlin, \&
  Tholen}]{farnocchia2014}
Farnocchia, D., Chesley, S., Chamberlin, A., \& Tholen, D. 2015, Icarus, 245,
  94

\bibitem[{{Farnocchia} {et~al.}(2013){Farnocchia}, {Chesley},
  {Vokrouhlick{\'y}}, {Milani}, {Spoto}, \& {Bottke}}]{farnocchia2013b}
{Farnocchia}, D., {Chesley}, S.~R., {Vokrouhlick{\'y}}, D., {et~al.} 2013,
  Icarus, 224, 1

\bibitem[{{Folkner} {et~al.}(2014){Folkner}, {Williams}, {Boggs}, {Park}, \&
  {Kuchynka}}]{folkner2014}
{Folkner}, W.~M., {Williams}, J.~G., {Boggs}, D.~H., {Park}, R.~S., \&
  {Kuchynka}, P. 2014, Interplanetary Network Progress Report, 196, 1

\bibitem[{{Fowler} \& {Chillemi}(1992)}]{fowler1992}
{Fowler}, J.~W., \& {Chillemi}, J.~R. 1992, The IRAS Minor Planet Survey, 17

\bibitem[{{Gehrels} {et~al.}(1970){Gehrels}, {Roemer}, {Taylor}, \&
  {Zellner}}]{gehrels1970}
{Gehrels}, T., {Roemer}, E., {Taylor}, R.~C., \& {Zellner}, B.~H. 1970, AJ, 75,
  186

\bibitem[{{Giorgini} {et~al.}(2002){Giorgini}, {Ostro}, {Benner}, {Chodas},
  {Chesley}, {Hudson}, {Nolan}, {Klemola}, {Standish}, {Jurgens}, {Rose},
  {Chamberlin}, {Yeomans}, \& {Margot}}]{giorgini2002}
{Giorgini}, J.~D., {Ostro}, S.~J., {Benner}, L.~A.~M., {et~al.} 2002, Science,
  296, 132

\bibitem[{{Goldstein}(1969)}]{goldstein1969}
{Goldstein}, R.~M. 1969, Icarus, 10, 430

\bibitem[{{Greenberg} \& {Margot}(2015)}]{greenberg2015}
{Greenberg}, A.~H., \& {Margot}, J.-L. 2015, \aj, 150, 114

\bibitem[{{Hagfors}(1964)}]{hagfors1964}
{Hagfors}, T. 1964, \jgr, 69, 3779

\bibitem[{{Harris}(1998)}]{harris1998}
{Harris}, A.~W. 1998, Icarus, 131, 291

\bibitem[{{Hudson} \& {Ostro}(1994)}]{hudson1994}
{Hudson}, R.~S., \& {Ostro}, S.~J. 1994, Science, 263, 940

\bibitem[{{Jewitt} \& {Li}(2010)}]{jewitt2010}
{Jewitt}, D., \& {Li}, J. 2010, \aj, 140, 1519

\bibitem[{Kalman(1960)}]{kalman1960}
Kalman, R.~E. 1960, Transactions of the ASME--Journal of Basic Engineering, 82,
  35

\bibitem[{{Magri} {et~al.}(2001){Magri}, {Consolmagno}, {Ostro}, {Benner}, \&
  {Beeney}}]{magr01}
{Magri}, C., {Consolmagno}, G.~J., {Ostro}, S.~J., {Benner}, L.~A.~M., \&
  {Beeney}, B.~R. 2001, Meteoritics and Planetary Science, 36, 1697

\bibitem[{{Magri} {et~al.}(2007{\natexlab{a}}){Magri}, {Nolan}, {Ostro}, \&
  {Giorgini}}]{magr07mbas}
{Magri}, C., {Nolan}, M.~C., {Ostro}, S.~J., \& {Giorgini}, J.~D.
  2007{\natexlab{a}}, Icarus, 186, 126

\bibitem[{{Magri} {et~al.}(2007{\natexlab{b}}){Magri}, {Ostro}, {Scheeres},
  {Nolan}, {Giorgini}, {Benner}, \& {Margot}}]{magri2007}
{Magri}, C., {Ostro}, S.~J., {Scheeres}, D.~J., {et~al.} 2007{\natexlab{b}},
  Icarus, 186, 152

\bibitem[{Magri {et~al.}(1999)Magri, Ostro, Rosema, Thomas, Mitchell, Campbell,
  Chandler, Shapiro, Giorgini, \& Yeomans}]{magr99}
Magri, C., Ostro, S., Rosema, K., {et~al.} 1999, Icarus, 140, 379

\bibitem[{{Mahapatra} {et~al.}(1999){Mahapatra}, {Ostro}, {Benner}, {Rosema},
  {Jurgens}, {Winkler}, {Rose}, {Giorgini}, {Yeomans}, \&
  {Slade}}]{mahapatra1999}
{Mahapatra}, P.~R., {Ostro}, S.~J., {Benner}, L.~A.~M., {et~al.} 1999, Planet.
  Space Sci., 47, 987

\bibitem[{{Mainzer} {et~al.}(2012){Mainzer}, {Grav}, {Masiero}, {Bauer},
  {Cutri}, {McMillan}, {Nugent}, {Tholen}, {Walker}, \& {Wright}}]{mainzer2012}
{Mainzer}, A., {Grav}, T., {Masiero}, J., {et~al.} 2012, \apjl, 760, L12

\bibitem[{Mandel(1964)}]{greenbook}
Mandel, J. 1964, The Statistical Analysis of Experimental Data (Dover
  Publications)

\bibitem[{{Margot} \& {Giorgini}(2010)}]{marg09iau261}
{Margot}, J.~L., \& {Giorgini}, J.~D. 2010, in IAU Symposium, Vol. 261, IAU
  Symposium, ed. {S.~A.~Klioner, P.~K.~Seidelmann, \& M.~H.~Soffel}, 183--188

\bibitem[{{Minor Planet Center}(2015)}]{mpcdb}
{Minor Planet Center}. 2015, Minor Planet database,
  \url{http://minorplanetcenter.net/db\_search}

\bibitem[{Nugent {et~al.}(2012)Nugent, Margot, Chesley, \&
  Vokrouhlick{\'y}}]{nuge12yark}
Nugent, C.~R., Margot, J.~L., Chesley, S.~R., \& Vokrouhlick{\'y}, D. 2012,
  Astronomical Journal, 144, 60

\bibitem[{{Ostro}(1993)}]{ostro1993b}
{Ostro}, S.~J. 1993, Reviews of Modern Physics, 65, 1235

\bibitem[{{Peterson}(1976)}]{peterson1976}
{Peterson}, C.~A. 1976, PhD thesis, Massachusetts Institute of Technology.

\bibitem[{{Pettengill} {et~al.}(1969){Pettengill}, {Shapiro}, {Ash}, {Ingalls},
  {Rainville}, {Smith}, \& {Stone}}]{pettengill1969}
{Pettengill}, G.~H., {Shapiro}, I.~I., {Ash}, M.~E., {et~al.} 1969, Icarus, 10,
  432

\bibitem[{Press {et~al.}(1992)Press, Teukolsky, Vetterling, \&
  Flannery}]{numericalrecipes}
Press, W.~H., Teukolsky, S.~A., Vetterling, W.~T., \& Flannery, B.~P. 1992,
  Numerical Recipes in C (2nd Ed.): The Art of Scientific Computing (New York,
  NY, USA: Cambridge University Press)

\bibitem[{{Rubincam}(2000)}]{rubincam2000}
{Rubincam}, D.~P. 2000, Icarus, 148, 2

\bibitem[{{Scheirich} {et~al.}(2015){Scheirich}, {Pravec}, {Jacobson}, {{\v
  D}urech}, {Ku{\v s}nir{\'a}k}, {Hornoch}, {Mottola}, {Mommert}, {Hellmich},
  {Pray}, {Polishook}, {Krugly}, {Inasaridze}, {Kvaratskhelia}, {Ayvazian},
  {Slyusarev}, {Pittichov{\'a}}, {Jehin}, {Manfroid}, {Gillon}, {Gal{\'a}d},
  {Pollock}, {Licandro}, {Al{\'{\i}}-Lagoa}, {Brinsfield}, \&
  {Molotov}}]{sche15}
{Scheirich}, P., {Pravec}, P., {Jacobson}, S.~A., {et~al.} 2015, Icarus, 245,
  56

\bibitem[{{Thomas} {et~al.}(2011){Thomas}, {Trilling}, {Emery}, {Mueller},
  {Hora}, {Benner}, {Bhattacharya}, {Bottke}, {Chesley}, {Delb{\'o}}, {Fazio},
  {Harris}, {Mainzer}, {Mommert}, {Morbidelli}, {Penprase}, {Smith}, {Spahr},
  \& {Stansberry}}]{thomas2011}
{Thomas}, C.~A., {Trilling}, D.~E., {Emery}, J.~P., {et~al.} 2011, \aj, 142, 85

\bibitem[{Verma \& Margot(2016)}]{verma2016}
Verma, A.~K., \& Margot, J.~L. 2016, JGR

\bibitem[{{Vokrouhlick{\'y}} \& {Farinella}(1998)}]{farinella1998}
{Vokrouhlick{\'y}}, D., \& {Farinella}, P. 1998, \aj, 116, 2032

\bibitem[{{Vokrouhlick{\'y}} {et~al.}(2000){Vokrouhlick{\'y}}, {Milani}, \&
  {Chesley}}]{vokrouhlicky2000}
{Vokrouhlick{\'y}}, D., {Milani}, A., \& {Chesley}, S.~R. 2000, Icarus, 148,
  118

\bibitem[{{Warner}(2015)}]{mpbulletin}
{Warner}, B.~D. 2015, Minor Planet Bulletin, Vol. 42,
  http://www.minorplanet.info/MPB/MPB\_42\-4.pdf

\end{thebibliography}
\end{document}